\documentclass[sigconf]{acmart}
\usepackage{graphicx}

\usepackage{ifthen}
\newboolean{techreport}
\usepackage{caption}
\usepackage{multirow}
\usepackage{subcaption}
\usepackage{hyperref}
\usepackage{amsmath}
\usepackage{tikz}
\usepackage[]{footmisc}
\usepackage{graphicx}
\usepackage{pifont}

\usepackage[ruled,linesnumbered,vlined]{algorithm2e}
\let\oldnl\nl
\newcommand{\nonl}{\renewcommand{\nl}{\let\nl\oldnl}}
\usepackage{tikz}

\definecolor{codegreen}{rgb}{0,0.6,0}
\definecolor{codegray}{rgb}{0.5,0.5,0.5}
\definecolor{codepurple}{HTML}{C42043}
\definecolor{backcolour}{rgb}{1,1,1}

\newcommand\numberstyle[1]{%
    \footnotesize
    \color{codegray}%
    \ttfamily
    \ifnum#1<10 0\fi#1 |%
}

\usepackage{enumitem}

\renewcommand\footnotetextcopyrightpermission[1]{} 

\newcommand\render{lumination }
\newcommand\renders{luminations }
\newcommand\renderNOSP{lumination}
\newcommand\rendersNOSP{luminations}

\begin{document}
\setboolean{techreport}{true}

\title{Holodeck:  Immersive 3D Displays Using Swarms of Flying Light Specks}
\subtitle{Extended Version}
\titlenote{A shorter version of this paper appeared in ACM-Multimedia Asia 2021~\cite{shahram2021}.}


\author{Shahram Ghandeharizadeh}
\affiliation{%
  \institution{University of Southern California}
  \city{Los Angeles}
  \state{California}
  \country{USA}
}
\email{shahram@usc.edu}

\begin{abstract}
Unmanned Aerial Vehicles (UAVs) have moved beyond a platform for hobbyists to enable environmental monitoring, journalism, film industry, search and rescue, package delivery, and entertainment.  This paper describes 3D displays using swarms of flying light specks, FLSs.  An FLS is a small (hundreds of micrometers in size) UAV with one or more light sources to generate different colors and textures with adjustable brightness.  A synchronized swarm of FLSs renders an illumination in a pre-specified 3D volume, an FLS display.  An FLS display provides true depth, enabling a user to perceive a scene more completely by analyzing its illumination from different angles.

An FLS display may either be non-immersive or immersive.  Both will support 3D acoustics.  Non-immersive FLS displays may be the size of a 1980's computer monitor, enabling a surgical team to observe and control micro robots performing heart surgery inside a patient's body.  Immersive FLS displays may be the size of a room, enabling users to interact with objects, e.g., a rock, a teapot.  An object with behavior will be constructed using FLS-matters.  FLS-matter will enable a user to touch and manipulate an object, e.g., a user may pick up a teapot or throw a rock.  An immersive and interactive FLS display will approximate Star Trek's Holodeck.  

A successful realization of the research ideas presented in this paper will provide fundamental insights into implementing a Holodeck using swarms of FLSs.  A Holodeck will transform the future of human communication and perception, and how we interact with information and data.  It will revolutionize the future of how we work, learn, play and entertain, receive medical care, and socialize.
\end{abstract}

\maketitle
\pagestyle{plain} 

\section{Introduction}\label{sec:intro}

This brave new ideas paper describes a next generation multimedia display realized using {\em Flying Light Specks}, FLSs.  An FLS is a miniature (hundreds of micrometer in size) Unmanned Aerial Vehicle (UAV) or\footnote{This paper uses the terms UAV and drone synonymously.} drone with either a reflective surface, one or more light sources, or both.  The intensity and brightness of both the light sources and the reflective surface may be adjusted to display different colors and textures.  A display may consist of a large swarm (millions, billions, and trillions) of FLSs that render a three dimensional (3D) depth scene that may include motion and 3D acoustics.  We use the term {\em illuminations}, {\em \renders} for short, to refer to such renderings and prevent confusion with today's 2D images, e.g., JPEG~\cite{hudson18}, and motion graphics, e.g., MPEG~\cite{mpeg}.

FLS displays will complement today's flat screens and may incorporate them for certain application use cases; see the gaming application of Section~\ref{sec:applications}.  Advantages of FLS displays compared with today's flat screens will be multi-fold.  First, their rendered \renders will capture reality more accurately.  A \render will be rendered by controlling position of FLSs in a 3D volume that has depth.  Second, their 3D \renders provide depth to the naked eye.  Moreover, a user looking at an FLS display from different angles may observe different content.  Third, FLS displays may be immersive by occupying an arbitrary space, transforming a wall, a ceiling, a room, a concert hall into a 3D depth display.  


With immersive displays, we introduce FLS-matter, a swarm of FLSs that will realize a geometric shape in 2D (e.g., square) and 3D (e.g., a pyramid).
Its lumination will scale to render its shape in different sizes.
FLS-matter is the building block of an object with behavior.
As detailed in Section~\ref{sec:flsmatter}, it will enable a user to interact with objects rendered as luminations.
An interactive immersive FLS display will manipulate both the user's depth perception and spatial awareness, approximating Star Trek's Holodeck.

\begin{table}[h]
  \caption{Number of FLSs for different sized spheres.}
  \label{tab:numFLSs}
  \begin{tabular}{lrr}
    \toprule
    Sphere (radius in mm) & Surface area & Number of FLSs \\
    \midrule
    Ping-pong (20) & 5,027 & 19,040 \\
    Tennis (34.29) & 14,776 & 55,968\\
    Basketball (120) & 180,956 & 685,438 \\
    Room size (1524) & 
    29,186,351
    & 110,554,359 \\
    \bottomrule
  \end{tabular}
\end{table}

To maintain the same resolution, the number of FLSs required by a \render increases as a function of its size.
Assuming today's 0.264 millimeter pixels, Table~\ref{tab:numFLSs} shows the number of FLSs required to display a sphere as a starting point for a 3D Earth.  As size is increased from a ping-pong to a basketball, the number of FLSs increases to hundreds of thousands.  With a room sized (10 feet by 10 feet) lumination, the number of FLSs is hundreds of millions. 


\subsection{Applications}\label{sec:applications}
It is tempting to motivate FLS displays in the context of a visionary system such as Memex~\cite{Bush45}.
We side-step this temptation and motivate FLS displays using several short term application areas that would benefit from them if they were available today.

Health-care:  FLS displays will be ideal interfaces for today's 3-D scanners such as the Magnetic Resonance Imaging~\cite{mri1999} (MRI) and surgical devices such as the da Vinci Robot~\cite{byrn2007}.  Today's MRI scanners show 3D volumes as a sequence of 2D images.  They are ripe to use a true 3D depth FLS display.  A typical da Vinci deployment provides its surgeon with a 3D video (using color depth) but not the assistants, even though they must collaborate.  There have been studies that quantify the benefits of providing assistants with their own 3D video monitor using color depth~\cite{davinci2014}.  A large FLS display will enable the surgeon and assistants to view the same 3D \render to collaborate effectively.  


Entertainment and gaming:  FLS displays will transform entertainment and gaming applications.  An example is multi-player fight games such as Mortal Kombat, Street Fighter, Fortnite and others.  FLS displays enable players to view the fight scene from different angles.  These \renders might be on a game board that sits on a table.  It may incorporate today's flat screens as its bottom. For example, in a scene that shows multiple characters standing on a wooden flooring, a flat screen will render the wooden flooring while FLSs delineate the 3D fighting characters.  Players may participate either remotely or in-person by either sitting or standing around the game board.  Players may use either a traditional game controller or wear body sensors that translate their movement into \rendersNOSP.  With the latter, when the players are in the same room, they may see the posture of their opponent(s) and anticipate their next move, introducing new gaming experiences.  Players may also get a good cardiovascular exercise from playing such games.

Design and manufacturing:  Display of a 3D design using flat 2D displays is not natural and may result in human errors.  3D printing empowers a designer to evaluate a final design prior to manufacturing it.  An FLS display will fill the gap between design and 3D print by enabling a designer to see their 3D designs and examine them from different perspectives in real-time.  With this application, an FLS display maybe the size of a game board that renders small designs (knots and bolts of an airplane seat) or the size of a room for larger designs (the airplane).  Both will enable a designer to zoom in and out to scrutinize components.  This may be done using a traditional keyboard and mouse, a gesture based interface, or more novel interfaces (see  Section~\ref{sec:novelinterfaces}).  FLS displays may be an integral part of design.  For example, they may render \renders of a simulation that stresses a design to analyze its failure thresholds.  This may include lumination renderings from real environments with manufactured components, e.g., flight recorder data from a crashed airplane.  These concepts apply to manufacturing where an FLS display may render the machine tools on a factory floor and how they manufacture a product, e.g., an automobile. 


\subsection{Contributions}
The contributions of this paper are:
\begin{enumerate}
    \item The idea of 3D displays using swarms of FLSs.
    \item A variant of the first idea, immersive and interactive displays using FLS-matter, approximating Star Trek's Holodeck.
    \item Requirement of FLS displays formulated as research challenges, including a survey of the relevant literature.
\end{enumerate}
To the best of our knowledge, the first two ideas are novel and not described elsewhere.  See~\cite{Alghamdi2021ArchitectureCA,Shakhatreh2019UnmannedAV,Chung2018ASO} for a survey of applications using UAVs and swarms.

The rest of this paper is organized as follows.  Section~\ref{sec:related} describes related work.  Sections~\ref{sec:display}-\ref{sec:standards} detail challenges of FLS displays.
Section~\ref{sec:discuss} concludes this paper.

\section{Related Work}\label{sec:related}
Our idea, a 3D display using swarm of FLSs, is inspired by today's drones used in light shows at night.  These outdoor and indoor shows are performed using illuminated, synchronized, and choreographed groups of drones arranged into various aerial formations. Almost any image and motion graphic can be recreated using a computer program that controls flight path and lighting of drones.  An FLS display is similar because each FLS is a drone and a \render is realized by synchronizing FLSs.  

With outdoor shows, a drone's GPS enables it to be at the right position and follow the correct path in the night sky to illuminate its light for the desired effect.  Millimeter drones of an FLS display must be positioned with micrometer precision.  Even if GPS was available, it would be too coarse~\cite{leong2016}.  The best some GPS devices claim is 3 meter accuracy 95\% of the time.  The United States government currently claims 7.8 meter with 95\% confidence for horizontal accuracy for civilian (SPS) GPS.  Real-Time Kinematic (RTK) GPS provides centimeter-level accuracy using additional base stations.  Vertical accuracy is worse with GPS solutions.  Thus, an FLS display requires novel positioning systems as described in Section~\ref{sec:positioning}.

There are other significant differences between show drones and FLS displays.  First, the number of drones in FLS displays will be significantly larger. 
According to the Guinness World Record, the most unmanned aerial vehicles in simultaneous flight for a light show is 3,281 for tens of minutes spread over a kilometre in Shanghai, China on March 29, 2021~\cite{guinessWorldRecord}.
The number of drones in an FLS display will be several orders of magnitude larger. 
Second, the area used by an FLS display may be significantly smaller. 
An FLS display may be as small as a standard game board.  This is orders of magnitude smaller than the night sky of outdoor drone shows and concert halls~\cite{dandrea2015} of indoor shows.  Third, each FLS is significantly smaller than show drones.  This enables FLSs to operate in tightly constrained environments in tight formations with higher accelerations, providing higher stability and more rapid response to changes in luminations~\cite{opticalpositioning1}. 

Systems such as Northrop Grumman’s Virtual Immersive Portable Environment (VIPE) Holodeck surrounds a user with 2D monitors~\cite{vipe2014}.  VIPE uses Kinect to detect a user action such as crawling and jumping to display the appropriate training content on the 2D monitors.  Our proposed immersive 3D FLS display may replace VIPE's 2D monitors, manipulating both the trainee's depth perception and spatial awareness to provide a more realistic experience.

Virtual Reality (VR)~\cite{Sutherland1968AHT,VRfisher,VR21} is a computer generated simulation of a 3D image or environment.  A user may interact with this environment using equipment such as a head-mounted display with a screen inside or gloves fitted with sensors.  Augmented Reality (AR)~\cite{Sutherland1968AHT,AR97,AR20} uses technology such as a mobile phone, tablet, or AR glasses to present the user with an enhanced version of the real physical world.  The enhancements may include digital visual elements, sound, or other sensory stimuli.   An immersive FLS display is different than both by providing the human sense of touch with no gloves and true depth with no mobile phone or head mounts. 

Fast 3D printing has been suggested as a mechanism to realize 3D displays~\cite{t1000}.  The idea is for a 3D object to arise out of a puddle in real-time to show a 3D image, and move and shift shape (by growing parts) around the display to show motion graphics.  This idea is inspired by the T-1000 robot from the Terminator 2 movie.  To realize such a display, 3D printing must become orders of magnitude faster and polymer chemistry has been suggested as a solution~\cite{t1000}.  An FLS display 
employs small drones with one or more light/reflective sources to generate color and texture.  
At the same time, an FLS display may incorporate fast 3D printing to represent objects in a \render that are stationary enough to match the speed of 3D printing.  This hybrid display requires lighting of the 3D printed objects.  Alternative solutions may include backlighting or FLSs that project light onto these printed objects.  The rest of this paper focuses on challenges of FLS displays and defers hybrids using 3D printing to future work.

\subsection{Recording of \renders}\label{sec:record}
Similar to today's cameras and camcorders that record a scene in 2D that is subsequently displayed using today’s monitors and TVs, 3D depth sensing cameras and camcorders will record \rendersNOSP.  We use the term {\em depthcam} to refer to these devices.

Today's depthcams can be classified into those that use either Time-of-Flight (ToF), Triangulation, or both.
For example, Microsoft's Kinect One uses a flash of light to illuminate a scene and thousands of sensors measure the return time from different parts of the scene.  
This pixelated ToF device measures depth of a million points per second.
A similar technology is employed by iPhone X and its dot projector~\cite{dotprojector2020} to create a 3D map of the user’s facial structure~\cite{appleP2017,appleP2020}.
The LiDAR Intel RealSense depthcam uses a variant of the same technique by employing Lidar used in some automobiles.
This depthcam uses a proprietary MEMS micro scanning technology~\cite{realsense2020} to acquire depth data for indoor applications.

The D4xx series of Intel RealSense~\cite{realsense2020} depthcams use triangulation technique via stereoscopic imaging.
It uses two cameras to record a scene and compare image patches from one camera with image patches in the other camera to collect depth data.
The basic idea is that those objects that are closer to the camera are displaced more in the horizontal axis.  
A limitation of this technique is with low texture scenes that cause the image patches from the different cameras to become ambiguous, e.g., a white wall with no texture.  
In this scenario, a ToF technique using infrared (IR) light may be used to acquire depth data.  This technique is used by Intel RealSense D43x and D45x depthcams.

A depthcam may include Inertial Measurement Unit (IMU) to capture 6 
degrees of freedom data\footnote{Six degrees of freedom refers to the freedom of movement of a depthcam 
in three-dimensional space: Forward/back, up/down, left/right, pitch, 
yaw, and roll.}, e.g., Intel RealSense D435i.
The structure from motion of the depthcam may be used to capture depth data~\cite{slam2003}.
Moreover, this data may be fused with data from other sensors to refine depth data and make it more accurate.



Today's depthcam technologies are used in games, to authenticate users, and to navigate a robot in an obstacle course.  We foresee their evolution as recording devices for FLS displays.




\section{Continuous Display of Motion \renders}\label{sec:display}

A rendering of a motion \render must be continuous and respect the spatio-temporal requirements of its motion.
This means FLSs must be at the right (x,y,z) coordinate of the display as a function of time and illuminate either (a) light at the pre-specified color, brightness, and texture or (b) provide a reflective surface for an external light source to generate color and texture.

\begin{figure}[!ht]
    \centering
    \includegraphics[width=1.0\linewidth]{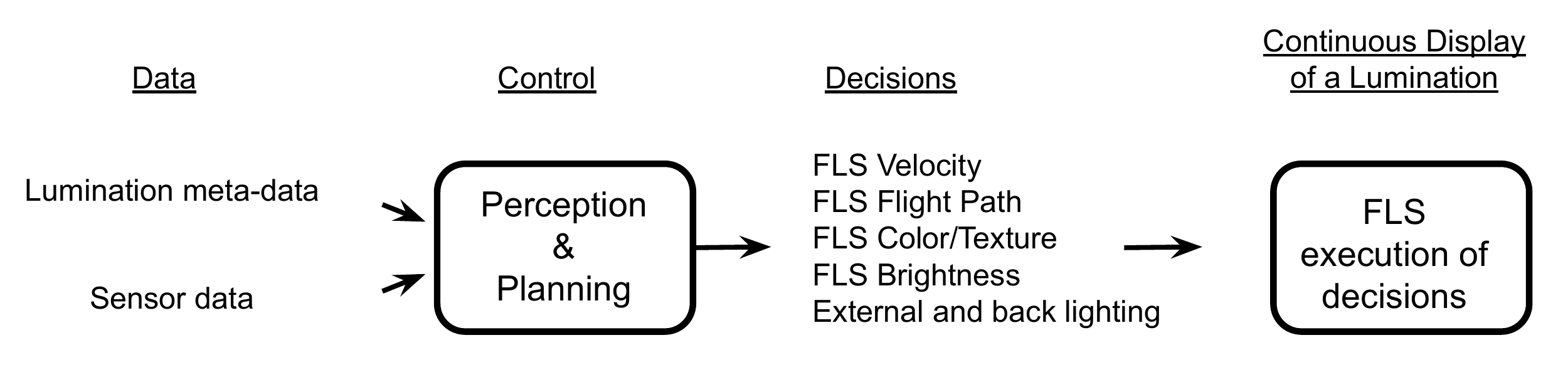}
    \caption{Continuous display using stream processing.}
    \label{fig:displayblocks}
\end{figure}

Figure~\ref{fig:displayblocks} shows the block diagram for continuous display adapted from the autonomous vehicles studies~\cite{Plathottam2018NextGD,uvaReview}.  Perception will transform streams of sensory data and the meta-data describing a lumination to a contextual understanding of the environment around each FLS and its goal.  Planning refers to the decisions that must be made by each FLS to execute its goal from a given start position.  It will generate streams of decisions that dictate the velocity, flight path, color/texture and brightness of light generated by each FLS.  These streams may also control external and back lighting of a scene.  A continuous display of a \render is the ability of FLSs to execute the collision-free flight decisions in a timely manner.

To illustrate, consider a \render of a running person.  If the pipeline of Figure~\ref{fig:displayblocks} is processed too slow then 
the display may show walking instead of running.  At its worst, the rendering may result in a distorted collection of lights in a 3D space.  A continuous display avoids these undesirable \renders by requiring each FLS or grouping of FLSs to be at the right coordinate at the right time to render the right color.


Continuous display is challenging for several reasons.  First, flight of FLSs must avoid collisions.  This challenge is exacerbated by complex scenes that may require a large number of FLSs that must render their color in close proximity to one another.  This is because no positioning system guarantees 100\% accuracy and reported position of FLSs may be erroneous.  Second, the scene to be displayed may not be known in advance.  An example is a real-time stream generated by a RealSense depthcam.  Third, failure of FLSs is the norm rather than the exception, see Section~\ref{sec:failure}.  A continuous display technique must render a lumination even though its FLSs are failing frequently.  Fourth, battery lifetime of FLSs is finite and power is scarce.  Fifth, the network bandwidth is finite and there is a latency for one FLS to communicate with another.

These challenges motivate algorithms and frameworks in support of a positioning system, a planner to compute flight path and lighting of each FLS as a function of time and space, a communication network for FLSs, and a system to charge FLS batteries.
Below, we describe these components in turn.




\subsection{Positioning system}\label{sec:positioning}

An indoor positioning system is required for an FLS to position itself at the right coordinate to render a motion lumination.  Even with a static lumination that has no motion, a positioning system is required to prevent FLSs from unpredictable drifting~\cite{preiss2017} that may cause crashes.  Outdoor drones use the global positioning system, GPS, to remain stationary.  GPS may not work indoors due to lack of line of sight with orbiting satellites\footnote{Even if GPSs worked indoors, its accuracy is too coarse for an FLS display, see Section~\ref{sec:related}.}.  A 3D FLS display requires a fine-grained positioning system to enable FLSs to maintain and control their position with hundreds of micrometer accuracy and 95\% confidence.  Indoor Location Systems (ILS) have been proposed based on Magnetic signals, Vision-based, Audible sound, Infrared (IR), Ultra-sound (UPS), Radio Frequency (RF) including RFID and WLAN, Bluetooth, Ultra-Wideband (UWB) and Ultra High Frequency (UHF), or a hybrid of these technologies.  See~\cite{zafari2019,liudarabi07,luca2014,joon2007,gusurvey2009} for a survey of these technologies including their strengths and limitations.  

Several studies report optical motion capture systems enable millimeter-level position tracking using multiple infrared cameras~\cite{opticalpositioning1,opticalpositioning2,preiss2017}.  This solution requires a unique marker arrangement for each drone.  The small size of an FLS may challenge the applicability of this solution.  For example,~\cite{preiss2017} reports that it was impossible to form unique markers for 49 drones with each measuring 92 millimeters diagonally.  They develop a method initialized with known positions that is subsequently updated using frame-by-frame tracking.  While such techniques may be a good starting point, we believe decentralized positioning systems that include active participation of FLSs will be developed.  In its simplest, FLSs may detect and either repel one another~\cite{repel1,repel2} or maintain a fixed geometry.  This will minimize the likelihood of FLS crashes when a positioning system provides inaccurate positions to FLSs.  

\subsection{Perception and Planning}
A planner computes flight path and lighting of each FLS as a function of time and space.  It may maneuver multiple FLSs in close proximity of one another to render a static \renderNOSP.

Several studies
detail path planning algorithms for UAVs, e.g., outdoor light shows~\cite{Sun2020PathPF}, search and rescue~\cite{rescueplanning2020}, services in an urban area~\cite{urbanplanning2021}, and environmental sensing~\cite{pathplanning2020}.
A challenge is to move from a given initial positions to a set of predefined targets while avoiding collisions with obstacles as well as other UAVs ~\cite{collisionfree2012,collisionfree2015,collisionavoidance2018,ReactiveCollisionAvoidance2008,ReactiveCollisionAvoidance2011,ReactiveCollisionAvoidance20112,reactiveColAvoidance2013,downwash1,downwash3,dcad2019,Engelhardt2016FlatnessbasedCF,navigation2017,reactiveColAvMorgan,reactiveColAvBaca,reactiveColAvMorganJ,speedAdjust2021,gameCollisionAvoidance2020,gameCollisionAvoidance2017,preiss2017,Ferrera2018Decentralized3C,planning2019,preiss2017whitewash,opticalpositioning1}.  Some studies consider {\em downwash}, a region of instability caused by the flight of one UAV that adversely impacts other UAVs entering this region~\cite{downwash1,dcad2019,preiss2017,downwash3,Ferrera2018Decentralized3C,planning2019}, e.g., loss of control or unpredictable behavior.  This effect is considered in collision avoidance by modelling each UAV as axis-aligned ellipsoids~\cite{dcad2019,preiss2017,downwash3} or cylinders~\cite{Ferrera2018Decentralized3C,planning2019} that results in a larger separation along the Z-axis.  The latter is appropriate for quadrotors. 

Today's collision avoidance techniques may be categorized into centralized~\cite{Sun2020PathPF,collisionavoidance2018,speedAdjust2021,gameCollisionAvoidance2017,collisionfree2015,gameCollisionAvoidance2020,preiss2017,opticalpositioning1,collisionfree2012,preiss2017whitewash,gameCollisionAvoidance2017} and decentralized~\cite{downwash3,dcad2019,reactiveColAvBaca,navigation2017,Ferrera2018Decentralized3C,reactiveColAvMorgan,ReactiveCollisionAvoidance2008,ReactiveCollisionAvoidance2011,downwash1,ReactiveCollisionAvoidance20112,reactiveColAvoidance2013,reactiveColAvMorganJ,speedAdjust2021,planning2019}.
A centralized algorithm executes on a CPU with abundant amount of memory and storage.  A decentralized algorithm may use the limited processing and storage capability of each FLS, requiring multiple of them to collaborate by communicating with one another.  Centralized algorithms may further be divided into online~\cite{collisionfree2015,gameCollisionAvoidance2020,preiss2017} or offline~\cite{Sun2020PathPF,collisionavoidance2018,speedAdjust2021,gameCollisionAvoidance2017,opticalpositioning1,collisionfree2012,preiss2017whitewash,gameCollisionAvoidance2017}.
Offline techniques assume a static environment with known obstacles, a known scene, and reliable UAVs.
Online techniques minimize the number of such assumptions.


To illustrate,~\cite{Sun2020PathPF} presents a centralized, offline algorithm to compute a flight/lighting plan for outdoor light show performances.  This algorithm requires drones to be placed in a field and in a specific arrangement.  Each drone is provided with a pre-computed flight path and timed light display.  All drones take off in a specific order, fly to the performance area, spread out and turn into the first performance pattern, change the formation into the next pattern, and repeat until all pre-specified performance patterns are complete.   Finally, the drones return into a formation and fly back to the takeoff position.  This algorithm is centralized because it assumes complete information of the field, intended display patterns and their duration, obstacles, and participating drones.  It is offline because it executes once, producing a plan for each drone and communicating it to that drone.  If during the light show, a new obstacle is introduced (e.g., a bird), an online algorithm is required to compute a new plan for the impacted drone(s).  This online algorithm is more complex than the offline algorithm because it must monitor the environment, detect new obstacles and the impacted drones, compute a new plan with a new flight path for each drone, identify the drones whose flight paths have changed, and communicate the new flight path to these drones.  An offline algorithm does not have these steps.  It may not even require drones to communicate because the drones are provided with a flight path once at the beginning.

It is important to note that centralized offline algorithms are of value because they may guarantee optimality in terms of metrics such as minimum flight time, time to realize a lumination, or amount of power consumed.  They can be used as yardsticks to evaluate online algorithms (both centralized and decentralized) that are typically heuristic based.

Collision avoidance techniques may further be categorized into those that evaluate their technique using a simulation~\cite{collisionavoidance2018,ReactiveCollisionAvoidance2008,dcad2019,reactiveColAvoidance2013,navigation2017,reactiveColAvMorgan,gameCollisionAvoidance2017,preiss2017}, an implementation using either outdoor UAVs, robots, or indoor quadrotors~\cite{collisionfree2015,collisionfree2012,speedAdjust2021}, or both~\cite{Sun2020PathPF,preiss2017whitewash,opticalpositioning1,downwash1,downwash3,ReactiveCollisionAvoidance2011,ReactiveCollisionAvoidance20112,reactiveColAvBaca,reactiveColAvMorganJ,gameCollisionAvoidance2020,Ferrera2018Decentralized3C,planning2019}.

Existing techniques and algorithms are a great starting point to develop an online planning component in support of continuous display.  
An ideal algorithm or framework should provide a scalable execution time as a function of computing resources.  This means a centralized algorithm should execute faster as a function of the number of CPU/GPU cores used to execute the algorithm.  Similarly, a decentralized algorithm should execute faster as a function of the number of participating FLSs.  Both concepts are challenging to realize in practice.

The concept of scalability extends beyond algorithms to include the display itself.  For example, a scalable positioning system will enable players of a game to stretch their 13 inch game board to 40 inches.  To maintain the resolution of luminations, users may add new FLSs.  It is interesting to note that with a decentralized algorithm for planning, the additional FLSs will increase the processing capability of the system to compute spatio-temporal plans automatically.  A centralized algorithm may require increased processing capability (e.g., additional cores) to schedule the new FLSs.  This may require a shut-down and restart of a game board.  An elastic decentralized algorithm will incorporate the new FLSs and continue the game without starving for processing capability.

\subsection{Communication}
FLSs must communicate to either receive a plan from a centralized planner or collaborate to compute a plan using a decentralized planner.  The communication will almost certainly be wireless.  The emerging 5 GHz 802.11n and 802.11ac offering real world bandwidths ranging from hundreds of megabits per second to more than 1 gigabit per second are promising.  This is because a spatio-temporal plan consists of meta-data that describes positions as a function of time with lighting specifications.  This data can be compressed.  Moreover, a plan may apply to a group of FLSs and broadcasted to all FLSs once. 

With a very large number of FLS, it is possible that FLSs exhaust the available wireless bandwidth.  There are networking technologies such as the power efficient ZigBee (802.15.4, 2.4 GHz) that provide for mesh communication between FLSs that are in close proximity of one another.  It may motivate use of multi-hop ad-hoc networks such as CAN~\cite{can2001} and Chord~\cite{chord2001} that route information to an FLS in a decentralized manner.
ZigBee is designed primarily for low-mobility devices such as those found in a home, e.g., smart bulbs and power outlets.
It is a great starting point to implement a network with mobility standards for FLS displays.
This network will enable FLSs flying in a tight formation (see FLS-Matter of Section~\ref{sec:flsmatter}) to communicate locally while traveling at high speeds.  This local communication should be independent of the global communication, providing for a large number of independent communications that do not interfere with one another.  In essence, the number of independent communications should scale as a function of the display size assuming fix sized FLSs.


\subsection{Power Conscious Display}


Continuous display algorithms must minimize the amount of power consumed to render a \renderNOSP.
This maximizes battery life of the participating FLSs. 
For example, a \render of a scene may consist of stationary objects that serve as its background, e.g., a building or a tree.  Depending on the duration of the scene, the stationary objects may be realized using FLSs that are docked into one another either horizontally and/or vertically.  These FLSs may use the base or sides of a display for support to stop flying altogether to save power.  The base or sides of the display may serve as a power source to charge the batteries of these FLSs.  These structures may serve as charging stations for other FLSs in their close vicinity.  

The emerging wireless chargers by companies such as Aeterlink, Motorola, Xiaomi, Oppo and others~\cite{nytimes2018} may be used to charge FLS batteries.  During CES 2021, Motorola demonstrated a charging technology that charged several phones at up to 1 meter away~\cite{motorola2021}.  Aeterlink's Airplug claims to power devices up to 20 meters away~\cite{lifewire2021,washingtonpost2021}.  These may be deployed in FLS game boards and non-immersive displays to charge FLS batteries continuously. A challenge is for this technology is to charge a large number of FLSs simultaneously.  

Wireless charging technology may not be appropriate for immersive FLS displays as wireless chargers emit EMF radiation which has been shown to be harmful to the human body~\cite{emf2018}.
This motivates a framework that recycles FLSs continuously.  This framework will replenish those FLSs with exhausted batteries with new ones.  It will charge their batteries and recycle them as new.
Note that this framework must be in the background and not interfere with continuous display of motion luminations.

With a motion \renderNOSP, an algorithm may pre-stage dark FLSs that light up to facilitate continuous motion.  Pre-staging FLSs will consume power.  However, its use may be more power efficient than flying FLSs around to render the lumination, e.g., by reducing the total distance travelled by FLSs. 



There will be a tradeoff between the quality of a \render and its consumed power.  A display may sacrifice the quality of a \render by using fewer FLSs, skipping positioning of certain FLSs all together, and controlling the brightness of light produced by FLSs.  These tradeoffs will have thresholds.  For example, a bright \render will consume more power and will most likely result in a higher quality display up to a certain threshold.  Beyond this threshold, increasing brightness will only consume power without providing a tangible human perceived improvement in quality.  This perception will most likely be impacted by environmental factors external to a display, e.g., lightning of the room.  Hence, an algorithms must identify the relevant environmental factors and incorporate them when trading power for quality of a \renderNOSP.  

A technique may save power by not rendering those portions of a lumination that are not visible to a user.  It may employ an eye tracker to compute the user's angle of view and identify regions of a 3D lumination that are hidden from the user.  An algorithm will use this information to minimize the number of FLSs with no impact on the user perceived quality of \renderNOSP, saving power.

As FLS displays are constructed, additional factors will almost certainly be identified that impact the tradeoff between power and quality of display.

\section{FLS-Matter}\label{sec:flsmatter}
FLS-matter is a pre-specified swarm of FLSs for 2D (e.g., square) and 3D (e.g., a pyramid) geometric shapes.
It is the building block of objects with behavior. 
Example objects include a rock, water, a teapot, a sword, etc.
The behavior associated with an object is inherited by its FLS-matters.  This behavior may further be specialized at the granularity of each FLS-matter.

FLS-matter detects human touch (force) when it is perturbed by having its FLSs pushed out of their expected position.
The default behavior will be for the FLS-matter to stop rendering its lumination, i.e., its FLSs go dark and scatter. 
An object may over-ride this behavior.
For example, it may model gravity to disintegrate its FLS-matters more naturally by having them fall to the bottom of the display prior to scattering and going dark.  

An interesting research direction is how to model a swarm adjusting its flight pattern to exert force back at the human touch without breaking FLSs or causing injury to the user. 
This is more natural, requiring the human to exert enough force (specified as a threshold by the object behavior) to justify an object breaking.
It requires models that quantify the amount of force exerted by a swarm based on its speed, flight pattern, mass of its FLSs, and the number of FLSs used at the perturbation point. For example, if an object is a rock and the user picks it up (exerts sufficient force against gravity) then the lumination must move up. 
If the user stops exerting force then the lumination must fall down at a speed dictated by the gravity pull and object mass.
Upon impact to the display floor, different objects may behave differently. For example, with a rock, if the impact force exceeds a threshold then its lumination may fragment into smaller luminations (rock pieces).
These are a few example behaviors assigned to an object and inherited by its FLS-matter(s).
Linden Lab's Second Life provides many example objects and avatars in its market place~\cite{secondlife}.
An immersive FLS display will be able to render these objects and avatars, providing users with an interactive 3D experience that is significantly richer than today's 2D monitors and screens of mobile devices.

\section{3D Acoustics}\label{sec:acoustic}
It is important for FLS displays to address acoustics from the start instead of an afterthought.
FLS displays require noise reduction~\cite{supress2012} to minimize and possibly eliminate the impact of unwanted noise attributed to their mechanical components, e.g., the buzzing sound of their rotors.
In addition, they require sound synthesis, propagation, and rendering techniques.
These techniques enhance the sense of realism and presence~\cite{better2002,soundSurvey2020} for FLS displays.
This is especially true for immersive FLS displays that require
spatialized audio rendering.  We describe each in turn.

Synthesis generates sound for an object realized using FLS-matter.
This may be aerodynamic sound such as that produced by an FLS-matter sword when waved in the air~\cite{aero2003}, blades of a fan or wind turbine rotating~\cite{Xu2020PhysicsGuidedSS}, and instruments that produce sound through air vibration~\cite{aero2015,printone2016}.
Other examples~\cite{soundSurvey2020} include flame sound~\cite{fire2010,fire2011,IHME20091545} and
water bubble sound~\cite{bubble2005,bubble2013}.

Propagation describes what a human ear perceives as sound transmitted from its source travels and changes due to factors such as scattering, reflection, refraction, diffraction and others.
A recent survey~\cite{soundSurvey2020} categorizes 
propagation techniques into wave-based~\cite{FAIRWEATHER2003759,mehra2013,raghu2018,directional2020,wild2017}, geometric~\cite{prop2012,KROKSTAD1968118,geometric95,geometry1985,4658194,LEWERS1993161}, diffraction~\cite{diffract2014,diffract1957,1451581,diffract2001}, and their hybrids~\cite{Granier1996ExperimentalAO,hybrid2013,hybrid2018}.
We describe a wave-based propagation technique in Section~\ref{sec:wavebased}.  For an overview of the other techniques see~\cite{soundSurvey2020}.

Rendering is the pipeline to display complex scenes with hundreds of sounds~\cite{soundRendering2004,soundParametric2014,Jot1991DigitalDN}.
Sounds may be associated with moving objects~\cite{rendering2004}.
A pipeline will include devices to deliver sound to a user, e.g., speakers, ear phones, etc.
With gaming, an FLS displays will almost certainly use an audio middleware such as FMOD.
This middleware adds sound or music to a game by providing APIs that receive events from the game engine.
It provides a workflow independent of the visuals that is integrated with the visuals~\cite{soundSurvey2020}.

\subsection{Wave-based sound propagation}\label{sec:wavebased}
This section presents a wave-based propagation technique~\cite{mehra2013,raghu2018} for immersive FLS displays.
Subsequently, it describes the role of machine learning~\cite{soundLearn2020} and generative adversarial networks~\cite{gans14} to expedite this technique.

An object-oriented approach to audio wave sound propagation models each object with an acoustic behavior characterized by a function~\cite{mehra2013,raghu2018}.  This function maps an arbitrary incident field on the object to a scattered field.  Inter-object relationships map the outgoing scatter field of one object to the incident field on an inter-connected object.  An FLS display will compute the acoustic response of a lumination to a sound source by solving a global linear system that considers inter-object wave propagation.  It performs a summation for all outgoing sources across all objects at the listener location for each ear.  The listener may wear ear phones to hear the 3D stereoscopic acoustics.

A recent study approximates the acoustic scatting field of an object using neural networks for interactive sound propagation in dynamic scenes~\cite{soundLearn2020}.  This technique requires an approximate training dataset and long training time.  Moreover, it has no error bounds on arbitrary objects.  

A hybrid approach that combines both the wave sound propagation and the neural network learning approach may be more suitable for FLS displays.
The idea is to use generative adversarial network~\cite{gans14}, GANs, to simultaneously train two models:  a generative model G that captures the data distribution and a discriminative model D that estimates the probability that a sample came from the training data.   The training procedures for G is to maximize the probability of D making a mistake.  G can be thought of as producing incorrect sound at the listener location.  D can be thought of as detecting the incorrect sound.  Training of the models is complete once G recovers the training data distribution and D equals to $\frac{1}{2}$ everywhere.  The strength of this approach is that the object oriented approach to audio wave sound propagation may generate training data specific to an application.  As the training database grows in size, the adversarial network trains itself to produce accurate results.  Representation of this data to capture the key elements of an application will have a significant impact on the correctness of the network.  It will constitute a main focus of FLS displays.

\section{Failures}\label{sec:failure}

FLSs will fail and must be garbage collected.  Some failed FLSs will recover and become operational after sometime similar to today's commodity servers.  An FLS display must detect these and recycle them as operational.

FLSs will fail for several reasons.  First, each FLS is a special-purpose computing system (similar to a cell phone) that may reboot.  Its reboot may cause it to fall to the base of a display and break.  Second, each FLS is a mechanical device with a mean time to failure, MTTF.  Given a large number of FLSs and assuming independent failures, the likelihood of {\em an} FLS failing is proportional to the number of FLSs\footnote{This is similar to a disk failing frequently in a thousand disk RAID even though each disk is highly reliable, i.e., has a high MTTF~\cite{gibson}.}.  Third, the positioning system (see Section~\ref{sec:positioning}) will most likely not be 100\% accurate.  Similar to today's GPS, it will provide positions with some confidence.  When reported positions are erroneous, FLSs may crash into one another or the bottom of the display and break.  Fourth, the heuristics that compute the fly path for individual FLSs may not be 100\% accurate, causing FLSs crashes for complex scenes.  These broken FLSs must be garbage collected.  A display may include redundant FLSs that substitute for the broken FLSs.  Overtime, a display may exhaust its redundant FLSs.  At that point, it must either be replaced with a new display or repaired by removing its garbage FLSs and replenishing it with new operational FLSs.  

It is possible for an FLS that rebooted to fall without breaking.  In this case, once its recovers, it will most likely identify itself as a functional FLS to the display and be re-incorporated. 

An interesting research direction will be to design robust algorithms that use fewer FLSs than required to provide the highest quality \renderNOSP.
Another is for a failed FLS to repel other FLSs as it falls to the ground along an arbitrary path.
A solution will enable FLSs to move away from the failed FLS in different contexts, e.g., rendering a complex scene, FLS-matter, etc.

An FLS display is comparable to today's data centers with a large number of servers.
Some of the errors and failures seen in data centers may manifest themselves in an FLS display.
It is important to identify lessons learnt from these errors and incorporate them in FLS displays to enhance their reliability.

\section{Display of missing data}\label{sec:missingdata}
A file containing a \render may lack the 3D depth, color, texture, and brightness of FLSs from all possible view angles.  With an FLS display, a user looking at a \render from one of these angles with no data may observe empty gaps.  A simple solution is to fill the gaps with FLSs that transition color from the foreground object to the background objects gradually.  Such a rendering must consider the volume occupied by each object.  Its limitation is that it may produce un-realistic illuminations.  For example, with a picture of people standing in front of a car, there is a clear gap between people and the car.  Filling the gap to look continuous is not realistic.  It will look more un-realistic as the distance between the people and the car increases.

An alternative is to infer the 3D geometry and structure of objects in the scene and render them to fill the empty gaps in a \renderNOSP.  There exist techniques that infer 3D geometry from one or multiple 2D images~\cite{VisualHull1994,VolumetricMethod96,ComputerVision2003,3DdeepLearning2015,nipsLearning3Dobjects,quanzeng2020}.
See~\cite{SurveyDeepLearning3Drecognition} for a survey of deep learning techniques.  A key challenge with today's techniques is that they must estimate depth from the 2D images.  This estimation will be significantly more accurate with a depthcam (see Section~\ref{sec:record}).  Instead of using 2D images, an FLS display will use a database of objects and their 3D geometry to fill the empty gaps.  It may include a user in the loop, expanding the system's database with the missing objects that are personal to that user, e.g., a user's pet.

Certain applications may require a hybrid of solutions.  For example, with a medical application, it may be best to show the empty gaps to a physician to inform them of the missing data.  The physician may then request the FLS display to fill the gaps by inferring the 3D geometry and structure using a database of the human anatomy or a deep learning technique trained using this database. 

\section{Novel User Interfaces}\label{sec:novelinterfaces}
Applications of Section~\ref{sec:applications} highlight the use of keyboard and mouse, gaming consoles, and gesture control.  While it is important for FLS displays to be backward compatible with traditional user interfaces, they may enable novel user interfaces.  For example, a user wearing a haptic glove may reach into an FLS display with their hand and interact with objects that constitute a \renderNOSP.  This is similar to today's touch screen but in 3D.  This enables a surgeon to reach into an FLS display and separate human organs rendered as one \render into different \renders and examine each individually.  Immersive room-size (and stadium-size) FLS displays will enable multiple users wearing special outfits to interact with virtual objects and one another.  We believe applications of FLS displays will motivate novel user interfaces that are more intuitive, effective, and relevant than today's user interfaces~\cite{mirri2019}.

\section{Standards and Market Forces}\label{sec:standards}

Each FLS must cost less than 1 cent to realize an economically viable 3D display.  Consider a 3D display with one million FLSs.  At 1 cent per FLS, this display costs \$10,000.  Even with one hundred thousand FLSs, the display still costs \$1,000.  Still expensive given that we are not considering the cost associated with the other display components, e.g., positioning system, power, casing, etc.  Past trends in consumer electronic goods show a significant drop in costs as they are adopted widely.  We believe mass production of FLS displays will reduce the cost of each FLS to lower than 1 cent. 

Standards will play an important role as FLS displays are developed.  From a micro perspective, these standards will identify functionality implemented internal and external to a display.  These standards will enable an FLS display to inter-operate with diverse devices, e.g., gaming consoles by different manufacturers, a set-top box for 3D shows, a computer server, a mobile device such as a smart phone, etc.  They will detail formatting of a \render and interfaces for middle-wares to enable different vendors to develop software and content for FLS displays.  They will also ensure the display is backward compatible with today's 2D images and videos.  From a macro perspective, standards will produce displays with features not available in today's displays.  One example is scalable displays that may adjust their size.  A user with a small (15 inch front) display will be able to decrease or increase the display size (say to 13 or 24 inches), see Section~\ref{sec:display}.  


\section{Conclusions}\label{sec:discuss}
3D displays using swarms of FLSs is a transformative idea that will pioneer a new era in multimedia systems.  Immersive and interactive 3D displays will resemble Star Trek's Holodeck.  Their realization will revolutionize the future of how we work, learn, play and entertain, receive medical care, communicate, and socialize.

While the technical challenges of FLS displays may appear daunting, it is important to remember the humble beginnings of today's displays as televisions with cathode ray tubes and a mechanical scanning system invented in\footnote{The first electronic television with no mechanical scanning was invented in 1927.} 1907.  Computer monitors using CRT technology with a monochrome display were first introduced in 1973.  While LED display technology was developed in 1977, it was not readily available for purchase until about 30 years later with 
release of one in late 2009.  LCD displays~\footnote{The difference between a LED and a LCD display is the backlighting. The first LCD monitors used CCFL instead of LEDs to illuminate the screen.} were introduced in the mid 1990s and outsold CRT monitors for the first time in 2003.  LCD displays became dominant starting 2007.  

Non-immersive FLS displays may observe an evolution similar to today's monitors to become a consumer electronic good.  We are hopeful that their introduction and adoption will be faster.  

Immersive FLS displays raise many interesting research challenges, see discussion of FLS-matter in Section~\ref{sec:flsmatter}.  It must consider safety of users and may require a longer development, deployment, and adoption life cycle.

\section{Acknowledgments}
We thank ACM-Multimedia Asia's anonymous reviewers (Brave New Ideas track) for their valuable comments. 

\bibliographystyle{ACM-Reference-Format}
\bibliography{main}  


\begin{thebibliography}{115}


\ifx \showCODEN    \undefined \def \showCODEN     #1{\unskip}     \fi
\ifx \showDOI      \undefined \def \showDOI       #1{#1}\fi
\ifx \showISBNx    \undefined \def \showISBNx     #1{\unskip}     \fi
\ifx \showISBNxiii \undefined \def \showISBNxiii  #1{\unskip}     \fi
\ifx \showISSN     \undefined \def \showISSN      #1{\unskip}     \fi
\ifx \showLCCN     \undefined \def \showLCCN      #1{\unskip}     \fi
\ifx \shownote     \undefined \def \shownote      #1{#1}          \fi
\ifx \showarticletitle \undefined \def \showarticletitle #1{#1}   \fi
\ifx \showURL      \undefined \def \showURL       {\relax}        \fi
\providecommand\bibfield[2]{#2}
\providecommand\bibinfo[2]{#2}
\providecommand\natexlab[1]{#1}
\providecommand\showeprint[2][]{arXiv:#2}

\bibitem[\protect\citeauthoryear{Alghamdi, Munir, and La}{Alghamdi
  et~al\mbox{.}}{2021}]%
        {Alghamdi2021ArchitectureCA}
\bibfield{author}{\bibinfo{person}{Yousef Alghamdi}, \bibinfo{person}{Arslan
  Munir}, {and} \bibinfo{person}{H. La}.} \bibinfo{year}{2021}\natexlab{}.
\newblock \showarticletitle{{Architecture, Classification, and Applications of
  Contemporary Unmanned Aerial Vehicles}}.
\newblock \bibinfo{journal}{\emph{IEEE Consumer Electronics Magazine}}
  (\bibinfo{year}{2021}), \bibinfo{pages}{1--10}.
\newblock


\bibitem[\protect\citeauthoryear{Allen and Raghuvanshi}{Allen and
  Raghuvanshi}{2015}]%
        {aero2015}
\bibfield{author}{\bibinfo{person}{Andrew Allen} {and} \bibinfo{person}{Nikunj
  Raghuvanshi}.} \bibinfo{year}{2015}\natexlab{}.
\newblock \showarticletitle{{Aerophones in Flatland: Interactive Wave
  Simulation of Wind Instruments}}.
\newblock \bibinfo{journal}{\emph{ACM Trans. Graph.}} \bibinfo{volume}{34},
  \bibinfo{number}{4}, Article \bibinfo{articleno}{134} (\bibinfo{date}{July}
  \bibinfo{year}{2015}), \bibinfo{numpages}{11}~pages.
\newblock
\showISSN{0730-0301}
\urldef\tempurl%
\url{https://doi.org/10.1145/2767001}
\showDOI{\tempurl}


\bibitem[\protect\citeauthoryear{Antani, Chandak, Savioja, and Manocha}{Antani
  et~al\mbox{.}}{2012}]%
        {prop2012}
\bibfield{author}{\bibinfo{person}{Lakulish Antani}, \bibinfo{person}{Anish
  Chandak}, \bibinfo{person}{Lauri Savioja}, {and} \bibinfo{person}{Dinesh
  Manocha}.} \bibinfo{year}{2012}\natexlab{}.
\newblock \showarticletitle{{Interactive Sound Propagation Using Compact
  Acoustic Transfer Operators}}.
\newblock \bibinfo{journal}{\emph{ACM Trans. Graph.}} \bibinfo{volume}{31},
  \bibinfo{number}{1}, Article \bibinfo{articleno}{7} (\bibinfo{date}{Feb.}
  \bibinfo{year}{2012}), \bibinfo{numpages}{12}~pages.
\newblock
\showISSN{0730-0301}
\urldef\tempurl%
\url{https://doi.org/10.1145/2077341.2077348}
\showDOI{\tempurl}


\bibitem[\protect\citeauthoryear{Arul and Manocha}{Arul and Manocha}{2020}]%
        {dcad2019}
\bibfield{author}{\bibinfo{person}{Senthil~Hariharan Arul} {and}
  \bibinfo{person}{D. Manocha}.} \bibinfo{year}{2020}\natexlab{}.
\newblock \showarticletitle{{DCAD: Decentralized Collision Avoidance With
  Dynamics Constraints for Agile Quadrotor Swarms}}.
\newblock \bibinfo{journal}{\emph{IEEE Robotics and Automation Letters}}
  \bibinfo{volume}{5} (\bibinfo{year}{2020}), \bibinfo{pages}{1191--1198}.
\newblock
\urldef\tempurl%
\url{https://doi.org/10.1109/LRA.2020.2967281}
\showDOI{\tempurl}


\bibitem[\protect\citeauthoryear{Augugliaro, Schoellig, and
  D'Andrea}{Augugliaro et~al\mbox{.}}{2012}]%
        {collisionfree2012}
\bibfield{author}{\bibinfo{person}{Federico Augugliaro},
  \bibinfo{person}{Angela Schoellig}, {and} \bibinfo{person}{Raffaello
  D'Andrea}.} \bibinfo{year}{2012}\natexlab{}.
\newblock \showarticletitle{{Generation of Collision-Free Trajectories for a
  Quadrocopter Fleet: A Sequential Convex Programming Approach}}. In
  \bibinfo{booktitle}{\emph{Proceedings of the IEEE/RSJ International
  Conference on Intelligent Robots and Systems. IEEE/RSJ International
  Conference on Intelligent Robots and Systems}}. \bibinfo{pages}{1917--1922}.
\newblock
\showISBNx{978-1-4673-1737-5}
\urldef\tempurl%
\url{https://doi.org/10.1109/IROS.2012.6385823}
\showDOI{\tempurl}


\bibitem[\protect\citeauthoryear{Azabl, Abdurazag, Shaban, and Mohamed}{Azabl
  et~al\mbox{.}}{2018}]%
        {emf2018}
\bibfield{author}{\bibinfo{person}{Azab Azabl}, \bibinfo{person}{Khalat
  Abdurazag}, \bibinfo{person}{Ebrahim Shaban}, {and} \bibinfo{person}{Albasha
  Mohamed}.} \bibinfo{year}{2018}\natexlab{}.
\newblock \showarticletitle{{Electromagnetic Fields and Its Harmful Effects on
  the Male Reproductive System}}.
\newblock \bibinfo{journal}{\emph{AIS Bioscience and Bioengineering}}
  \bibinfo{volume}{4}, \bibinfo{number}{1} (\bibinfo{year}{2018}),
  \bibinfo{pages}{1--13}.
\newblock


\bibitem[\protect\citeauthoryear{Bareiss and van~den Berg}{Bareiss and van~den
  Berg}{2013}]%
        {downwash3}
\bibfield{author}{\bibinfo{person}{Daman Bareiss} {and} \bibinfo{person}{Joran
  van~den Berg}.} \bibinfo{year}{2013}\natexlab{}.
\newblock \showarticletitle{{Reciprocal Collision Avoidance for Robots with
  Linear Dynamics using LQR-Obstacles}}. In
  \bibinfo{booktitle}{\emph{Proceedings - IEEE International Conference on
  Robotics and Automation}}. \bibinfo{pages}{3847--3853}.
\newblock
\showISBNx{978-1-4673-5641-1}
\urldef\tempurl%
\url{https://doi.org/10.1109/ICRA.2013.6631118}
\showDOI{\tempurl}


\bibitem[\protect\citeauthoryear{Biot and Tolstoy}{Biot and Tolstoy}{1957}]%
        {diffract1957}
\bibfield{author}{\bibinfo{person}{M.~A. Biot} {and} \bibinfo{person}{I.
  Tolstoy}.} \bibinfo{year}{1957}\natexlab{}.
\newblock \showarticletitle{{Formulation of Wave Propagation in Infinite Media
  by Normal Coordinates with an Application to Diffraction}}.
\newblock \bibinfo{journal}{\emph{The Journal of the Acoustical Society of
  America}} \bibinfo{volume}{29}, \bibinfo{number}{3} (\bibinfo{year}{1957}),
  \bibinfo{pages}{381--391}.
\newblock
\urldef\tempurl%
\url{https://doi.org/10.1121/1.1908899}
\showDOI{\tempurl}
\showeprint{https://doi.org/10.1121/1.1908899}


\bibitem[\protect\citeauthoryear{Brodsky}{Brodsky}{5196}]%
        {lifewire2021}
\bibfield{author}{\bibinfo{person}{Sascha Brodsky}.} \bibinfo{year}{Lifewire,
  May 18, 2021,
  https://www.lifewire.com/your-smartphone-could-soon-charge-over-the-air-5185196}\natexlab{}.
\newblock \bibinfo{title}{{Your Smartphone Could Soon Charge Over the Air}}.
\newblock
\newblock


\bibitem[\protect\citeauthoryear{Brown}{Brown}{ air}]%
        {washingtonpost2021}
\bibfield{author}{\bibinfo{person}{Dalvin Brown}.} \bibinfo{year}{The
  Washington Post, March 5, 2021,
  https://www.washingtonpost.com/technology/2021/03/05/wireless-charging-over-the-air/}\natexlab{}.
\newblock \bibinfo{title}{{One Innovation We Won't be Seeing Soon: Over-The-Air
  Charging}}.
\newblock
\newblock


\bibitem[\protect\citeauthoryear{Bush}{Bush}{1945}]%
        {Bush45}
\bibfield{author}{\bibinfo{person}{Vannevar Bush}.}
  \bibinfo{year}{1945}\natexlab{}.
\newblock \showarticletitle{{As We May Think}}.
\newblock \bibinfo{journal}{\emph{Atlantic Monthly}}  \bibinfo{volume}{176}
  (\bibinfo{year}{1945}), \bibinfo{pages}{101--108}.
\newblock


\bibitem[\protect\citeauthoryear{Byrn, Schluender, Divino, Conrad, Gurland,
  Shlasko, and Szold}{Byrn et~al\mbox{.}}{2007}]%
        {byrn2007}
\bibfield{author}{\bibinfo{person}{John~C. Byrn}, \bibinfo{person}{Stefanie
  Schluender}, \bibinfo{person}{Celia~M. Divino}, \bibinfo{person}{John
  Conrad}, \bibinfo{person}{Brooke Gurland}, \bibinfo{person}{Edward Shlasko},
  {and} \bibinfo{person}{Amir Szold}.} \bibinfo{year}{2007}\natexlab{}.
\newblock \showarticletitle{{Three-Dimensional Imaging Improves Surgical
  Performance for both Novice and Experienced Operators using the da Vinci
  Robot System}}.
\newblock \bibinfo{journal}{\emph{The American Journal of Surgery}}
  \bibinfo{volume}{193}, \bibinfo{number}{4} (\bibinfo{year}{2007}),
  \bibinfo{pages}{519--522}.
\newblock
\showISSN{0002-9610}
\urldef\tempurl%
\url{https://doi.org/10.1016/j.amjsurg.2006.06.042}
\showDOI{\tempurl}


\bibitem[\protect\citeauthoryear{Báča, Hert, Loianno, Saska, and
  Kumar}{Báča et~al\mbox{.}}{2018}]%
        {reactiveColAvBaca}
\bibfield{author}{\bibinfo{person}{Tomáš Báča}, \bibinfo{person}{Daniel
  Hert}, \bibinfo{person}{Giuseppe Loianno}, \bibinfo{person}{Martin Saska},
  {and} \bibinfo{person}{Vijay Kumar}.} \bibinfo{year}{2018}\natexlab{}.
\newblock \showarticletitle{{Model Predictive Trajectory Tracking and Collision
  Avoidance for Reliable Outdoor Deployment of Unmanned Aerial Vehicles}}. In
  \bibinfo{booktitle}{\emph{IEEE/RSJ International Conference on Intelligent
  Robots and Systems (IROS)}}. \bibinfo{pages}{6753--6760}.
\newblock
\urldef\tempurl%
\url{https://doi.org/10.1109/IROS.2018.8594266}
\showDOI{\tempurl}


\bibitem[\protect\citeauthoryear{Campion, Ranganathan, and Faruque}{Campion
  et~al\mbox{.}}{2019}]%
        {uvaReview}
\bibfield{author}{\bibinfo{person}{Mitch Campion}, \bibinfo{person}{Prakash
  Ranganathan}, {and} \bibinfo{person}{Saleh Faruque}.}
  \bibinfo{year}{2019}\natexlab{}.
\newblock \showarticletitle{{UAV Swarm Communication and Control Architectures:
  A Review}}.
\newblock \bibinfo{journal}{\emph{Journal of Unmanned Vehicle Systems}}
  \bibinfo{volume}{7}, \bibinfo{number}{2} (\bibinfo{year}{2019}),
  \bibinfo{pages}{93--106}.
\newblock
\urldef\tempurl%
\url{https://doi.org/10.1139/juvs-2018-0009}
\showDOI{\tempurl}


\bibitem[\protect\citeauthoryear{Cappello, Garcin, Mao, Sassano, Paranjape, and
  Mylvaganam}{Cappello et~al\mbox{.}}{2020}]%
        {gameCollisionAvoidance2020}
\bibfield{author}{\bibinfo{person}{D. Cappello}, \bibinfo{person}{S. Garcin},
  \bibinfo{person}{Z. Mao}, \bibinfo{person}{M. Sassano}, \bibinfo{person}{A.
  Paranjape}, {and} \bibinfo{person}{T. Mylvaganam}.}
  \bibinfo{year}{2020}\natexlab{}.
\newblock \showarticletitle{{A Hybrid Controller for Multi-Agent Collision
  Avoidance via a Differential Game Formulation}}.
\newblock \bibinfo{journal}{\emph{IEEE Transactions on Control Systems
  Technology}}  \bibinfo{volume}{PP} (\bibinfo{date}{07} \bibinfo{year}{2020}),
  \bibinfo{pages}{1--8}.
\newblock
\urldef\tempurl%
\url{https://doi.org/10.1109/TCST.2020.3005602}
\showDOI{\tempurl}


\bibitem[\protect\citeauthoryear{Chadwick and James}{Chadwick and
  James}{2011}]%
        {fire2011}
\bibfield{author}{\bibinfo{person}{Jeffrey~N. Chadwick} {and}
  \bibinfo{person}{Doug~L. James}.} \bibinfo{year}{2011}\natexlab{}.
\newblock \showarticletitle{{Animating Fire with Sound}}.
\newblock \bibinfo{journal}{\emph{ACM Trans. Graph.}} \bibinfo{volume}{30},
  \bibinfo{number}{4}, Article \bibinfo{articleno}{84} (\bibinfo{date}{July}
  \bibinfo{year}{2011}), \bibinfo{numpages}{8}~pages.
\newblock
\showISSN{0730-0301}
\urldef\tempurl%
\url{https://doi.org/10.1145/2010324.1964979}
\showDOI{\tempurl}


\bibitem[\protect\citeauthoryear{Chaitanya, Raghuvanshi, Godin, Zhang,
  Nowrouzezahrai, and Snyder}{Chaitanya et~al\mbox{.}}{2020}]%
        {directional2020}
\bibfield{author}{\bibinfo{person}{Chakravarty R.~Alla Chaitanya},
  \bibinfo{person}{Nikunj Raghuvanshi}, \bibinfo{person}{Keith~W. Godin},
  \bibinfo{person}{Zechen Zhang}, \bibinfo{person}{Derek Nowrouzezahrai}, {and}
  \bibinfo{person}{John~M. Snyder}.} \bibinfo{year}{2020}\natexlab{}.
\newblock \showarticletitle{{Directional Sources and Listeners in Interactive
  Sound Propagation Using Reciprocal Wave Field Coding}}.
\newblock \bibinfo{journal}{\emph{ACM Trans. Graph.}} \bibinfo{volume}{39},
  \bibinfo{number}{4}, Article \bibinfo{articleno}{44} (\bibinfo{date}{July}
  \bibinfo{year}{2020}), \bibinfo{numpages}{14}~pages.
\newblock
\showISSN{0730-0301}
\urldef\tempurl%
\url{https://doi.org/10.1145/3386569.3392459}
\showDOI{\tempurl}


\bibitem[\protect\citeauthoryear{Chandak, Lauterbach, Taylor, Ren, and
  Manocha}{Chandak et~al\mbox{.}}{2008}]%
        {4658194}
\bibfield{author}{\bibinfo{person}{Anish Chandak}, \bibinfo{person}{Christian
  Lauterbach}, \bibinfo{person}{Micah Taylor}, \bibinfo{person}{Zhimin Ren},
  {and} \bibinfo{person}{Dinesh Manocha}.} \bibinfo{year}{2008}\natexlab{}.
\newblock \showarticletitle{{AD-Frustum: Adaptive Frustum Tracing for
  Interactive Sound Propagation}}.
\newblock \bibinfo{journal}{\emph{IEEE Transactions on Visualization and
  Computer Graphics}} \bibinfo{volume}{14}, \bibinfo{number}{6}
  (\bibinfo{year}{2008}), \bibinfo{pages}{1707--1722}.
\newblock
\urldef\tempurl%
\url{https://doi.org/10.1109/TVCG.2008.111}
\showDOI{\tempurl}


\bibitem[\protect\citeauthoryear{Chen, Cutler, and How}{Chen
  et~al\mbox{.}}{2015}]%
        {collisionfree2015}
\bibfield{author}{\bibinfo{person}{Yu~Fan Chen}, \bibinfo{person}{Mark Cutler},
  {and} \bibinfo{person}{Jonathan How}.} \bibinfo{year}{2015}\natexlab{}.
\newblock \showarticletitle{{Decoupled Multiagent Path Planning via Incremental
  Sequential Convex Programming}}.
\newblock \bibinfo{journal}{\emph{Proceedings - IEEE International Conference
  on Robotics and Automation}}  \bibinfo{volume}{2015} (\bibinfo{date}{06}
  \bibinfo{year}{2015}), \bibinfo{pages}{5954--5961}.
\newblock
\urldef\tempurl%
\url{https://doi.org/10.1109/ICRA.2015.7140034}
\showDOI{\tempurl}


\bibitem[\protect\citeauthoryear{Cheng, Zhu, Liu, Xu, and Lin}{Cheng
  et~al\mbox{.}}{2017}]%
        {navigation2017}
\bibfield{author}{\bibinfo{person}{Hui Cheng}, \bibinfo{person}{Q. Zhu},
  \bibinfo{person}{Z. Liu}, \bibinfo{person}{Tianye Xu}, {and}
  \bibinfo{person}{Liang Lin}.} \bibinfo{year}{2017}\natexlab{}.
\newblock \showarticletitle{{Decentralized Navigation of Multiple Agents Based
  on ORCA and Model Predictive Control}}.
\newblock \bibinfo{journal}{\emph{2017 IEEE/RSJ International Conference on
  Intelligent Robots and Systems (IROS)}} (\bibinfo{year}{2017}),
  \bibinfo{pages}{3446--3451}.
\newblock
\urldef\tempurl%
\url{https://doi.org/10.1109/IROS.2017.8206184}
\showDOI{\tempurl}


\bibitem[\protect\citeauthoryear{Chung, Paranjape, Dames, Shen, and
  Kumar}{Chung et~al\mbox{.}}{2018}]%
        {Chung2018ASO}
\bibfield{author}{\bibinfo{person}{Soon-Jo Chung}, \bibinfo{person}{A.
  Paranjape}, \bibinfo{person}{P. Dames}, \bibinfo{person}{S. Shen}, {and}
  \bibinfo{person}{Vijay~R. Kumar}.} \bibinfo{year}{2018}\natexlab{}.
\newblock \showarticletitle{{A Survey on Aerial Swarm Robotics}}.
\newblock \bibinfo{journal}{\emph{IEEE Transactions on Robotics}}
  \bibinfo{volume}{34} (\bibinfo{year}{2018}), \bibinfo{pages}{837--855}.
\newblock


\bibitem[\protect\citeauthoryear{Curless and Levoy}{Curless and Levoy}{1996}]%
        {VolumetricMethod96}
\bibfield{author}{\bibinfo{person}{Brian Curless} {and} \bibinfo{person}{Marc
  Levoy}.} \bibinfo{year}{1996}\natexlab{}.
\newblock \showarticletitle{{A Volumetric Method for Building Complex Models
  from Range Images}}. In \bibinfo{booktitle}{\emph{Proceedings of the 23rd
  Annual Conference on Computer Graphics and Interactive Techniques}}
  \emph{(\bibinfo{series}{SIGGRAPH '96})}. \bibinfo{publisher}{Association for
  Computing Machinery}, \bibinfo{address}{New York, NY, USA},
  \bibinfo{pages}{303–312}.
\newblock
\showISBNx{0897917464}
\urldef\tempurl%
\url{https://doi.org/10.1145/237170.237269}
\showDOI{\tempurl}


\bibitem[\protect\citeauthoryear{Dadoun, Kirkpatrick, and Walsh}{Dadoun
  et~al\mbox{.}}{1985}]%
        {geometry1985}
\bibfield{author}{\bibinfo{person}{Norm Dadoun}, \bibinfo{person}{David~G.
  Kirkpatrick}, {and} \bibinfo{person}{John~P. Walsh}.}
  \bibinfo{year}{1985}\natexlab{}.
\newblock \showarticletitle{{The Geometry of Beam Tracing}}. In
  \bibinfo{booktitle}{\emph{Proceedings of the First Annual Symposium on
  Computational Geometry}} (Baltimore, Maryland, USA)
  \emph{(\bibinfo{series}{SCG '85})}. \bibinfo{publisher}{Association for
  Computing Machinery}, \bibinfo{address}{New York, NY, USA},
  \bibinfo{pages}{55–61}.
\newblock
\showISBNx{0897911636}
\urldef\tempurl%
\url{https://doi.org/10.1145/323233.323241}
\showDOI{\tempurl}


\bibitem[\protect\citeauthoryear{Davison}{Davison}{2003}]%
        {slam2003}
\bibfield{author}{\bibinfo{person}{Andrew~J. Davison}.}
  \bibinfo{year}{2003}\natexlab{}.
\newblock \showarticletitle{{Real-Time Simultaneous Localisation and Mapping
  with a Single Camera}}. In \bibinfo{booktitle}{\emph{Proceedings of the Ninth
  IEEE International Conference on Computer Vision - Volume 2}}
  \emph{(\bibinfo{series}{ICCV '03})}. \bibinfo{publisher}{IEEE Computer
  Society}, \bibinfo{address}{USA}, \bibinfo{pages}{1403}.
\newblock
\showISBNx{0769519504}


\bibitem[\protect\citeauthoryear{Deane}{Deane}{2013}]%
        {bubble2013}
\bibfield{author}{\bibinfo{person}{Grant~B. Deane}.}
  \bibinfo{year}{2013}\natexlab{}.
\newblock \showarticletitle{{Determining the Bubble Cap Film Thickness of
  Bursting Bubbles from their Acoustic Emissions}}.
\newblock \bibinfo{journal}{\emph{The Journal of the Acoustical Society of
  America}}  \bibinfo{volume}{133 2} (\bibinfo{year}{2013}),
  \bibinfo{pages}{EL69--75}.
\newblock


\bibitem[\protect\citeauthoryear{DeSimone}{DeSimone}{0146}]%
        {t1000}
\bibfield{author}{\bibinfo{person}{Joseph DeSimone}.} \bibinfo{year}{Onstage at
  TED2015. See https://www.ted.com/
  talks/joseph\_desimone\_what\_\if\_3d\_printing\_was\_100x\_faster?language=en\#t-20146}\natexlab{}.
\newblock \bibinfo{title}{{What if 3D Printing was 100x Faster?}}
\newblock
\newblock


\bibitem[\protect\citeauthoryear{Dobashi, Yamamoto, and Nishita}{Dobashi
  et~al\mbox{.}}{2003}]%
        {aero2003}
\bibfield{author}{\bibinfo{person}{Yoshinori Dobashi},
  \bibinfo{person}{Tsuyoshi Yamamoto}, {and} \bibinfo{person}{Tomoyuki
  Nishita}.} \bibinfo{year}{2003}\natexlab{}.
\newblock \showarticletitle{{Real-Time Rendering of Aerodynamic Sound Using
  Sound Textures Based on Computational Fluid Dynamics}}.
\newblock \bibinfo{journal}{\emph{ACM Trans. Graph.}} \bibinfo{volume}{22},
  \bibinfo{number}{3} (\bibinfo{date}{July} \bibinfo{year}{2003}),
  \bibinfo{pages}{732–740}.
\newblock
\showISSN{0730-0301}
\urldef\tempurl%
\url{https://doi.org/10.1145/882262.882339}
\showDOI{\tempurl}


\bibitem[\protect\citeauthoryear{Doel}{Doel}{2005}]%
        {bubble2005}
\bibfield{author}{\bibinfo{person}{Kees van~den Doel}.}
  \bibinfo{year}{2005}\natexlab{}.
\newblock \showarticletitle{{Physically Based Models for Liquid Sounds}}.
\newblock \bibinfo{journal}{\emph{ACM Trans. Appl. Percept.}}
  \bibinfo{volume}{2}, \bibinfo{number}{4} (\bibinfo{date}{Oct.}
  \bibinfo{year}{2005}), \bibinfo{pages}{534–546}.
\newblock
\showISSN{1544-3558}
\urldef\tempurl%
\url{https://doi.org/10.1145/1101530.1101554}
\showDOI{\tempurl}


\bibitem[\protect\citeauthoryear{Du, Turner, Dzitsiuk, Prasso, Duarte,
  Dourgarian, Afonso, Pascoal, Gladstone, Cruces, Izadi, Kowdle, Tsotsos, and
  Kim}{Du et~al\mbox{.}}{2020}]%
        {AR20}
\bibfield{author}{\bibinfo{person}{Ruofei Du}, \bibinfo{person}{Eric Turner},
  \bibinfo{person}{Maksym Dzitsiuk}, \bibinfo{person}{Luca Prasso},
  \bibinfo{person}{Ivo Duarte}, \bibinfo{person}{Jason Dourgarian},
  \bibinfo{person}{Jo{\~{a}}o Afonso}, \bibinfo{person}{Jose Pascoal},
  \bibinfo{person}{Josh Gladstone}, \bibinfo{person}{Nuno Cruces},
  \bibinfo{person}{Shahram Izadi}, \bibinfo{person}{Adarsh Kowdle},
  \bibinfo{person}{Konstantine Tsotsos}, {and} \bibinfo{person}{David Kim}.}
  \bibinfo{year}{2020}\natexlab{}.
\newblock \showarticletitle{{DepthLab: Real-time 3D Interaction with Depth Maps
  for Mobile Augmented Reality}}. In \bibinfo{booktitle}{\emph{{UIST} '20: The
  33rd Annual {ACM} Symposium on User Interface Software and Technology,
  Virtual Event, USA, October 20-23, 2020}},
  \bibfield{editor}{\bibinfo{person}{Shamsi~T. Iqbal},
  \bibinfo{person}{Karon~E. MacLean}, \bibinfo{person}{Fanny Chevalier}, {and}
  \bibinfo{person}{Stefanie Mueller}} (Eds.). \bibinfo{publisher}{{ACM}},
  \bibinfo{pages}{829--843}.
\newblock
\urldef\tempurl%
\url{https://doi.org/10.1145/3379337.3415881}
\showDOI{\tempurl}


\bibitem[\protect\citeauthoryear{Engelhardt, Konrad, Schafer, and
  Abel}{Engelhardt et~al\mbox{.}}{2016}]%
        {Engelhardt2016FlatnessbasedCF}
\bibfield{author}{\bibinfo{person}{T. Engelhardt}, \bibinfo{person}{T. Konrad},
  \bibinfo{person}{Bjorn~E. Schafer}, {and} \bibinfo{person}{D. Abel}.}
  \bibinfo{year}{2016}\natexlab{}.
\newblock \showarticletitle{{Flatness-Based Control for a Quadrotor Camera
  Helicopter using Model Predictive Control Trajectory Generation}}.
\newblock \bibinfo{journal}{\emph{24th Mediterranean Conference on Control and
  Automation (MED)}} (\bibinfo{year}{2016}), \bibinfo{pages}{852--859}.
\newblock


\bibitem[\protect\citeauthoryear{Fairweather, Karageorghis, and
  Martin}{Fairweather et~al\mbox{.}}{2003}]%
        {FAIRWEATHER2003759}
\bibfield{author}{\bibinfo{person}{Graeme Fairweather},
  \bibinfo{person}{Andreas Karageorghis}, {and} \bibinfo{person}{P.A. Martin}.}
  \bibinfo{year}{2003}\natexlab{}.
\newblock \showarticletitle{{The Method of Fundamental Solutions for Scattering
  and Radiation Problems}}.
\newblock \bibinfo{journal}{\emph{Engineering Analysis with Boundary Elements}}
  \bibinfo{volume}{27}, \bibinfo{number}{7} (\bibinfo{year}{2003}),
  \bibinfo{pages}{759--769}.
\newblock
\showISSN{0955-7997}
\urldef\tempurl%
\url{https://doi.org/10.1016/S0955-7997(03)00017-1}
\showDOI{\tempurl}
\newblock
\shownote{Special issue on Acoustics.}


\bibitem[\protect\citeauthoryear{Feiner, Macintyre, Höllerer, and
  Webster}{Feiner et~al\mbox{.}}{1997}]%
        {AR97}
\bibfield{author}{\bibinfo{person}{Steven Feiner}, \bibinfo{person}{Blair
  Macintyre}, \bibinfo{person}{Tobias Höllerer}, {and}
  \bibinfo{person}{Anthony Webster}.} \bibinfo{year}{1997}\natexlab{}.
\newblock \showarticletitle{{A Touring Machine: Prototyping 3D Mobile Augmented
  Reality Systems for Exploring the Urban Environment}}.
\newblock \bibinfo{journal}{\emph{Personal Technologies}}  \bibinfo{volume}{1},
  \bibinfo{pages}{74--81}.
\newblock
\urldef\tempurl%
\url{https://doi.org/10.1007/BF01682023}
\showDOI{\tempurl}


\bibitem[\protect\citeauthoryear{Ferrera, Alc{\'a}ntara, Capit{\'a}n,
  Casta{\~n}o, Marr{\'o}n, and Ollero}{Ferrera et~al\mbox{.}}{2018}]%
        {Ferrera2018Decentralized3C}
\bibfield{author}{\bibinfo{person}{Eduardo Ferrera}, \bibinfo{person}{Alfonso
  Alc{\'a}ntara}, \bibinfo{person}{J. Capit{\'a}n}, \bibinfo{person}{{\'A}.~R.
  Casta{\~n}o}, \bibinfo{person}{P. Marr{\'o}n}, {and} \bibinfo{person}{A.
  Ollero}.} \bibinfo{year}{2018}\natexlab{}.
\newblock \showarticletitle{{Decentralized 3D Collision Avoidance for Multiple
  UAVs in Outdoor Environments}}.
\newblock \bibinfo{journal}{\emph{Sensors (Basel, Switzerland)}}
  \bibinfo{volume}{18} (\bibinfo{year}{2018}).
\newblock


\bibitem[\protect\citeauthoryear{Fisher}{Fisher}{2016}]%
        {VRfisher}
\bibfield{author}{\bibinfo{person}{Scott~S. Fisher}.}
  \bibinfo{year}{2016}\natexlab{}.
\newblock \showarticletitle{{The Nasa Ames Viewlab Project-a Brief History}}.
\newblock \bibinfo{journal}{\emph{Presence: Teleoper. Virtual Environ.}}
  \bibinfo{volume}{24}, \bibinfo{number}{4} (\bibinfo{date}{Dec.}
  \bibinfo{year}{2016}), \bibinfo{pages}{339–348}.
\newblock
\showISSN{1054-7460}
\urldef\tempurl%
\url{https://doi.org/10.1162/PRES_a_00277}
\showDOI{\tempurl}


\bibitem[\protect\citeauthoryear{Ghandeharizadeh}{Ghandeharizadeh}{2021}]%
        {shahram2021}
\bibfield{author}{\bibinfo{person}{Shahram Ghandeharizadeh}.}
  \bibinfo{year}{2021}\natexlab{}.
\newblock \showarticletitle{{Holodeck: Immersive 3D Displays Using Swarms of
  Flying Light Specks}}. In \bibinfo{booktitle}{\emph{ACM Multimedia Asia}}
  (Gold Coast, Australia).
\newblock
\urldef\tempurl%
\url{https://doi.org/10.1145/3469877.3493698}
\showDOI{\tempurl}


\bibitem[\protect\citeauthoryear{Goodfellow, Pouget{-}Abadie, Mirza, Xu,
  Warde{-}Farley, Ozair, Courville, and Bengio}{Goodfellow
  et~al\mbox{.}}{2014}]%
        {gans14}
\bibfield{author}{\bibinfo{person}{Ian~J. Goodfellow}, \bibinfo{person}{Jean
  Pouget{-}Abadie}, \bibinfo{person}{Mehdi Mirza}, \bibinfo{person}{Bing Xu},
  \bibinfo{person}{David Warde{-}Farley}, \bibinfo{person}{Sherjil Ozair},
  \bibinfo{person}{Aaron~C. Courville}, {and} \bibinfo{person}{Yoshua Bengio}.}
  \bibinfo{year}{2014}\natexlab{}.
\newblock \showarticletitle{{Generative Adversarial Nets}}. In
  \bibinfo{booktitle}{\emph{Advances in Neural Information Processing Systems
  27: Annual Conference on Neural Information Processing Systems 2014, December
  8-13 2014, Montreal, Quebec, Canada}},
  \bibfield{editor}{\bibinfo{person}{Zoubin Ghahramani}, \bibinfo{person}{Max
  Welling}, \bibinfo{person}{Corinna Cortes}, \bibinfo{person}{Neil~D.
  Lawrence}, {and} \bibinfo{person}{Kilian~Q. Weinberger}} (Eds.).
  \bibinfo{pages}{2672--2680}.
\newblock
\urldef\tempurl%
\url{https://proceedings.neurips.cc/paper/2014/hash/5ca3e9b122f61f8f06494c97b1afccf3-Abstract.html}
\showURL{%
\tempurl}


\bibitem[\protect\citeauthoryear{Granier, Kleiner, Dalenb{\"a}ck, and
  Svensson}{Granier et~al\mbox{.}}{1996}]%
        {Granier1996ExperimentalAO}
\bibfield{author}{\bibinfo{person}{Emmanuel Granier}, \bibinfo{person}{Mendel
  Kleiner}, \bibinfo{person}{Bengt-Inge Dalenb{\"a}ck}, {and}
  \bibinfo{person}{Peter Svensson}.} \bibinfo{year}{1996}\natexlab{}.
\newblock \showarticletitle{{Experimental Auralization of Car Audio
  Installations}}.
\newblock \bibinfo{journal}{\emph{Journal of The Audio Engineering Society}}
  \bibinfo{volume}{44} (\bibinfo{year}{1996}), \bibinfo{pages}{835--849}.
\newblock


\bibitem[\protect\citeauthoryear{Gu, Lo, and Niemegeers}{Gu
  et~al\mbox{.}}{2009}]%
        {gusurvey2009}
\bibfield{author}{\bibinfo{person}{Yanying Gu}, \bibinfo{person}{Anthony C.~C.
  Lo}, {and} \bibinfo{person}{Ignas~G. Niemegeers}.}
  \bibinfo{year}{2009}\natexlab{}.
\newblock \showarticletitle{A Survey of Indoor Positioning Systems for Wireless
  Personal Networks}.
\newblock \bibinfo{journal}{\emph{{IEEE} Commun. Surv. Tutorials}}
  \bibinfo{volume}{11}, \bibinfo{number}{1} (\bibinfo{year}{2009}),
  \bibinfo{pages}{13--32}.
\newblock
\urldef\tempurl%
\url{https://doi.org/10.1109/SURV.2009.090103}
\showDOI{\tempurl}


\bibitem[\protect\citeauthoryear{Hamer, Widmer, and Drandrea}{Hamer
  et~al\mbox{.}}{2018}]%
        {collisionavoidance2018}
\bibfield{author}{\bibinfo{person}{Michael Hamer}, \bibinfo{person}{Lino
  Widmer}, {and} \bibinfo{person}{Raffaello Drandrea}.}
  \bibinfo{year}{2018}\natexlab{}.
\newblock \showarticletitle{{Fast Generation of Collision-Free Trajectories for
  Robot Swarms Using GPU Acceleration}}.
\newblock \bibinfo{journal}{\emph{IEEE Access}}  \bibinfo{volume}{PP}
  (\bibinfo{date}{12} \bibinfo{year}{2018}), \bibinfo{pages}{1--1}.
\newblock
\urldef\tempurl%
\url{https://doi.org/10.1109/ACCESS.2018.2889533}
\showDOI{\tempurl}


\bibitem[\protect\citeauthoryear{Han, Laga, and Bennamoun}{Han
  et~al\mbox{.}}{2021}]%
        {SurveyDeepLearning3Drecognition}
\bibfield{author}{\bibinfo{person}{X. Han}, \bibinfo{person}{H. Laga}, {and}
  \bibinfo{person}{M. Bennamoun}.} \bibinfo{year}{2021}\natexlab{}.
\newblock \showarticletitle{{Image-Based 3D Object Reconstruction:
  State-of-the-Art and Trends in the Deep Learning Era}}.
\newblock \bibinfo{journal}{\emph{IEEE Transactions on Pattern Analysis \&
  Machine Intelligence}} \bibinfo{volume}{43}, \bibinfo{number}{05}
  (\bibinfo{date}{may} \bibinfo{year}{2021}), \bibinfo{pages}{1578--1604}.
\newblock
\showISSN{1939-3539}
\urldef\tempurl%
\url{https://doi.org/10.1109/TPAMI.2019.2954885}
\showDOI{\tempurl}


\bibitem[\protect\citeauthoryear{Hartley and Zisserman}{Hartley and
  Zisserman}{2003}]%
        {ComputerVision2003}
\bibfield{author}{\bibinfo{person}{Richard Hartley} {and}
  \bibinfo{person}{Andrew Zisserman}.} \bibinfo{year}{2003}\natexlab{}.
\newblock \bibinfo{booktitle}{\emph{{Multiple View Geometry in Computer
  Vision}} (\bibinfo{edition}{2} ed.)}.
\newblock \bibinfo{publisher}{Cambridge University Press},
  \bibinfo{address}{USA}.
\newblock
\showISBNx{0521540518}


\bibitem[\protect\citeauthoryear{Hayat, Yanmaz, Bettstetter, and Brown}{Hayat
  et~al\mbox{.}}{2020}]%
        {rescueplanning2020}
\bibfield{author}{\bibinfo{person}{Samira Hayat}, \bibinfo{person}{Evsen
  Yanmaz}, \bibinfo{person}{Christian Bettstetter}, {and}
  \bibinfo{person}{Timothy Brown}.} \bibinfo{year}{2020}\natexlab{}.
\newblock \showarticletitle{Multi-objective Drone Path Planning for Search and
  Rescue with Quality-of-Service Requirements}.
\newblock \bibinfo{journal}{\emph{Autonomous Robots}}  \bibinfo{volume}{44}
  (\bibinfo{date}{09} \bibinfo{year}{2020}).
\newblock
\urldef\tempurl%
\url{https://doi.org/10.1007/s10514-020-09926-9}
\showDOI{\tempurl}


\bibitem[\protect\citeauthoryear{Henderson, Liu, Folden, Tilmon, Jayasuriya,
  and Koppal}{Henderson et~al\mbox{.}}{2020}]%
        {dotprojector2020}
\bibfield{author}{\bibinfo{person}{Kristofer Henderson},
  \bibinfo{person}{Xiaomeng Liu}, \bibinfo{person}{Justin Folden},
  \bibinfo{person}{Brevin Tilmon}, \bibinfo{person}{Suren Jayasuriya}, {and}
  \bibinfo{person}{Sanjeev Koppal}.} \bibinfo{year}{2020}\natexlab{}.
\newblock \showarticletitle{{Design and Calibration of a Fast Flying-Dot
  Projector for Dynamic Light Transport Acquisition}}.
\newblock \bibinfo{journal}{\emph{IEEE Transactions on Computational Imaging}}
  \bibinfo{volume}{6} (\bibinfo{year}{2020}), \bibinfo{pages}{529--543}.
\newblock
\urldef\tempurl%
\url{https://doi.org/10.1109/TCI.2020.2964246}
\showDOI{\tempurl}


\bibitem[\protect\citeauthoryear{Heo, Pyeon, Kim, and Sohn}{Heo
  et~al\mbox{.}}{2007}]%
        {joon2007}
\bibfield{author}{\bibinfo{person}{Joon Heo}, \bibinfo{person}{Mu~Wook Pyeon},
  \bibinfo{person}{Jung~Whan Kim}, {and} \bibinfo{person}{Hong-Gyoo Sohn}.}
  \bibinfo{year}{2007}\natexlab{}.
\newblock \showarticletitle{Towards the Optimal Design of an RFID-Based
  Positioning System for the Ubiquitous Computing Environment}. In
  \bibinfo{booktitle}{\emph{Proceedings of the 2nd International Conference on
  Rough Sets and Knowledge Technology}} (Toronto, Canada)
  \emph{(\bibinfo{series}{RSKT'07})}. \bibinfo{publisher}{Springer-Verlag},
  \bibinfo{address}{Berlin, Heidelberg}, \bibinfo{pages}{331–338}.
\newblock
\showISBNx{9783540724575}


\bibitem[\protect\citeauthoryear{Herkes, Olsen, and Uellenberg}{Herkes
  et~al\mbox{.}}{2012}]%
        {supress2012}
\bibfield{author}{\bibinfo{person}{W. Herkes}, \bibinfo{person}{R. Olsen},
  {and} \bibinfo{person}{S. Uellenberg}.} \bibinfo{year}{2012}\natexlab{}.
\newblock \showarticletitle{{The Quiet Technology Demonstrator Program: Flight
  Validation of Airplane Noise-Reduction Concepts}}. In
  \bibinfo{booktitle}{\emph{12th AIAA/CEAS Aeroacoustics Conference (27th AIAA
  Aeroacoustics Conference)}}.
\newblock
\urldef\tempurl%
\url{https://doi.org/10.2514/6.2006-2720}
\showDOI{\tempurl}
\showeprint{https://arc.aiaa.org/doi/pdf/10.2514/6.2006-2720}


\bibitem[\protect\citeauthoryear{Hudson, L{\'{e}}ger, Niss, Sebesty{\'{e}}n,
  and Vaaben}{Hudson et~al\mbox{.}}{2018}]%
        {hudson18}
\bibfield{author}{\bibinfo{person}{Graham Hudson}, \bibinfo{person}{Alain
  L{\'{e}}ger}, \bibinfo{person}{Birger Niss}, \bibinfo{person}{Istvan
  Sebesty{\'{e}}n}, {and} \bibinfo{person}{J{\o}rgen Vaaben}.}
  \bibinfo{year}{2018}\natexlab{}.
\newblock \showarticletitle{{JPEG-1 Standard 25 Years: Past, Present, and
  Future Reasons for a Success}}.
\newblock \bibinfo{journal}{\emph{J. Electronic Imaging}} \bibinfo{volume}{27},
  \bibinfo{number}{04} (\bibinfo{year}{2018}).
\newblock
\urldef\tempurl%
\url{https://doi.org/10.1117/1.JEI.27.4.040901}
\showDOI{\tempurl}


\bibitem[\protect\citeauthoryear{Ihme, Pitsch, and Bodony}{Ihme
  et~al\mbox{.}}{2009}]%
        {IHME20091545}
\bibfield{author}{\bibinfo{person}{Matthias Ihme}, \bibinfo{person}{Heinz
  Pitsch}, {and} \bibinfo{person}{Daniel Bodony}.}
  \bibinfo{year}{2009}\natexlab{}.
\newblock \showarticletitle{{Radiation of Noise in Turbulent Non-Premixed
  Flames}}.
\newblock \bibinfo{journal}{\emph{Proceedings of the Combustion Institute}}
  \bibinfo{volume}{32}, \bibinfo{number}{1} (\bibinfo{year}{2009}),
  \bibinfo{pages}{1545--1553}.
\newblock
\showISSN{1540-7489}
\urldef\tempurl%
\url{https://doi.org/10.1016/j.proci.2008.06.137}
\showDOI{\tempurl}


\bibitem[\protect\citeauthoryear{ISO/IEC JTC 1/SC 29, SC 29/WG 11}{ISO/IEC JTC
  1/SC 29, SC 29/WG 11}{2001}]%
        {mpeg}
ISO/IEC JTC 1/SC 29, SC 29/WG 11 \bibinfo{year}{2001}\natexlab{}.
\newblock \bibinfo{booktitle}{\emph{{Coding of Moving Pictures and Audio}}}.
\newblock \bibinfo{type}{Standard}. \bibinfo{institution}{International
  Organization for Standardization}, \bibinfo{address}{Tokyo, Japan}.
\newblock


\bibitem[\protect\citeauthoryear{Jot and Chaigne}{Jot and Chaigne}{1991}]%
        {Jot1991DigitalDN}
\bibfield{author}{\bibinfo{person}{Jean-Marc Jot} {and}
  \bibinfo{person}{Antoine Chaigne}.} \bibinfo{year}{1991}\natexlab{}.
\newblock \showarticletitle{Digital Delay Networks for Designing Artificial
  Reverberators}.
\newblock \bibinfo{journal}{\emph{Journal of The Audio Engineering Society}}
  (\bibinfo{year}{1991}).
\newblock


\bibitem[\protect\citeauthoryear{Kouyoumjian and Pathak}{Kouyoumjian and
  Pathak}{1974}]%
        {1451581}
\bibfield{author}{\bibinfo{person}{R.G. Kouyoumjian} {and}
  \bibinfo{person}{P.H. Pathak}.} \bibinfo{year}{1974}\natexlab{}.
\newblock \showarticletitle{A Uniform Geometrical Theory of Diffraction for an
  Edge in a Perfectly Conducting Surface}.
\newblock \bibinfo{journal}{\emph{Proc. IEEE}} \bibinfo{volume}{62},
  \bibinfo{number}{11} (\bibinfo{year}{1974}), \bibinfo{pages}{1448--1461}.
\newblock
\urldef\tempurl%
\url{https://doi.org/10.1109/PROC.1974.9651}
\showDOI{\tempurl}


\bibitem[\protect\citeauthoryear{Krokstad, Strom, and Sørsdal}{Krokstad
  et~al\mbox{.}}{1968}]%
        {KROKSTAD1968118}
\bibfield{author}{\bibinfo{person}{A. Krokstad}, \bibinfo{person}{S. Strom},
  {and} \bibinfo{person}{S. Sørsdal}.} \bibinfo{year}{1968}\natexlab{}.
\newblock \showarticletitle{{Calculating the Acoustical Room Response by the
  Use of a Ray Tracing Technique}}.
\newblock \bibinfo{journal}{\emph{Journal of Sound and Vibration}}
  \bibinfo{volume}{8}, \bibinfo{number}{1} (\bibinfo{year}{1968}),
  \bibinfo{pages}{118--125}.
\newblock
\showISSN{0022-460X}
\urldef\tempurl%
\url{https://doi.org/10.1016/0022-460X(68)90198-3}
\showDOI{\tempurl}


\bibitem[\protect\citeauthoryear{Kugler}{Kugler}{2021}]%
        {VR21}
\bibfield{author}{\bibinfo{person}{Logan Kugler}.}
  \bibinfo{year}{2021}\natexlab{}.
\newblock \showarticletitle{{The State of Virtual Reality Hardware}}.
\newblock \bibinfo{journal}{\emph{Commun. ACM}} \bibinfo{volume}{64},
  \bibinfo{number}{2} (\bibinfo{date}{Jan.} \bibinfo{year}{2021}),
  \bibinfo{pages}{15–16}.
\newblock
\showISSN{0001-0782}
\urldef\tempurl%
\url{https://doi.org/10.1145/3441290}
\showDOI{\tempurl}


\bibitem[\protect\citeauthoryear{Kushleyev, Mellinger, Powers, and
  Kumar}{Kushleyev et~al\mbox{.}}{2013}]%
        {opticalpositioning1}
\bibfield{author}{\bibinfo{person}{Alex Kushleyev}, \bibinfo{person}{Daniel
  Mellinger}, \bibinfo{person}{Caitlin Powers}, {and} \bibinfo{person}{Vijay
  Kumar}.} \bibinfo{year}{2013}\natexlab{}.
\newblock \showarticletitle{{Towards a Swarm of Agile Micro Quadrotors}}.
\newblock \bibinfo{journal}{\emph{Autonomous Robots}}  \bibinfo{volume}{35}
  (\bibinfo{date}{11} \bibinfo{year}{2013}), \bibinfo{pages}{573--7527}.
\newblock
\urldef\tempurl%
\url{https://doi.org/10.1007/s10514-013-9349-9}
\showDOI{\tempurl}


\bibitem[\protect\citeauthoryear{Labs}{Labs}{2021}]%
        {secondlife}
\bibfield{author}{\bibinfo{person}{Linden Labs}.}
  \bibinfo{year}{2021}\natexlab{}.
\newblock \bibinfo{title}{{Second Life Marketplace, Oct 28, 2021. See
  https://marketplace.secondlife.com/}}.
\newblock
\newblock


\bibitem[\protect\citeauthoryear{Lalish}{Lalish}{2009}]%
        {repel2}
\bibfield{author}{\bibinfo{person}{Emmett Lalish}.}
  \bibinfo{year}{2009}\natexlab{}.
\newblock \bibinfo{title}{{Distributed Reactive Collision Avoidance, Ph.D.
  Disertation, University of Washington}}.
\newblock
\newblock


\bibitem[\protect\citeauthoryear{Larsson and Kleiner}{Larsson and
  Kleiner}{2002}]%
        {better2002}
\bibfield{author}{\bibinfo{person}{Pontus Larsson} {and}
  \bibinfo{person}{Mendel Kleiner}.} \bibinfo{year}{2002}\natexlab{}.
\newblock \showarticletitle{{Better Presence and Performance in Virtual
  Environments by Improved Binaural Sound Rendering}}. In
  \bibinfo{booktitle}{\emph{Virtual, Synthetic, and Entertainment Audio
  conference}}. \bibinfo{publisher}{AES}, \bibinfo{address}{Espoo, Finland}.
\newblock


\bibitem[\protect\citeauthoryear{Laurentini}{Laurentini}{1994}]%
        {VisualHull1994}
\bibfield{author}{\bibinfo{person}{A. Laurentini}.}
  \bibinfo{year}{1994}\natexlab{}.
\newblock \showarticletitle{{The Visual Hull Concept for Silhouette-Based Image
  Understanding}}.
\newblock \bibinfo{journal}{\emph{IEEE Trans. Pattern Anal. Mach. Intell.}}
  \bibinfo{volume}{16}, \bibinfo{number}{2} (\bibinfo{date}{Feb.}
  \bibinfo{year}{1994}), \bibinfo{pages}{150–162}.
\newblock
\showISSN{0162-8828}
\urldef\tempurl%
\url{https://doi.org/10.1109/34.273735}
\showDOI{\tempurl}


\bibitem[\protect\citeauthoryear{Lee, Jones, Ridenour, Bennett, Majors, Melito,
  and Wilson}{Lee et~al\mbox{.}}{2016}]%
        {leong2016}
\bibfield{author}{\bibinfo{person}{Leong Lee}, \bibinfo{person}{Matthew Jones},
  \bibinfo{person}{Gregory~S. Ridenour}, \bibinfo{person}{Sam~J. Bennett},
  \bibinfo{person}{Arisha~C. Majors}, \bibinfo{person}{Bianca~L. Melito}, {and}
  \bibinfo{person}{Michael~J. Wilson}.} \bibinfo{year}{2016}\natexlab{}.
\newblock \showarticletitle{{Comparison of Accuracy and Precision of
  GPS-Enabled Mobile Devices}}. In \bibinfo{booktitle}{\emph{2016 IEEE
  International Conference on Computer and Information Technology (CIT)}}.
  \bibinfo{publisher}{IEEE}, \bibinfo{pages}{73--82}.
\newblock
\urldef\tempurl%
\url{https://doi.org/10.1109/CIT.2016.94}
\showDOI{\tempurl}


\bibitem[\protect\citeauthoryear{Lewers}{Lewers}{1993}]%
        {LEWERS1993161}
\bibfield{author}{\bibinfo{person}{T. Lewers}.}
  \bibinfo{year}{1993}\natexlab{}.
\newblock \showarticletitle{{A Combined Beam Tracing and Radiant Exchange
  Computer Model of Room Acoustics}}.
\newblock \bibinfo{journal}{\emph{Applied Acoustics}} \bibinfo{volume}{38},
  \bibinfo{number}{2} (\bibinfo{year}{1993}), \bibinfo{pages}{161--178}.
\newblock
\showISSN{0003-682X}
\urldef\tempurl%
\url{https://doi.org/10.1016/0003-682X(93)90049-C}
\showDOI{\tempurl}


\bibitem[\protect\citeauthoryear{Liu, Darabi, Banerjee, and Liu}{Liu
  et~al\mbox{.}}{2007}]%
        {liudarabi07}
\bibfield{author}{\bibinfo{person}{Hui Liu}, \bibinfo{person}{Houshang Darabi},
  \bibinfo{person}{Pat~P. Banerjee}, {and} \bibinfo{person}{Jing Liu}.}
  \bibinfo{year}{2007}\natexlab{}.
\newblock \showarticletitle{Survey of Wireless Indoor Positioning Techniques
  and Systems}.
\newblock \bibinfo{journal}{\emph{{IEEE} Trans. Syst. Man Cybern. Part {C}}}
  \bibinfo{volume}{37}, \bibinfo{number}{6} (\bibinfo{year}{2007}),
  \bibinfo{pages}{1067--1080}.
\newblock
\urldef\tempurl%
\url{https://doi.org/10.1109/TSMCC.2007.905750}
\showDOI{\tempurl}


\bibitem[\protect\citeauthoryear{Liu and Manocha}{Liu and Manocha}{2020}]%
        {soundSurvey2020}
\bibfield{author}{\bibinfo{person}{Shiguang Liu} {and} \bibinfo{person}{Dinesh
  Manocha}.} \bibinfo{year}{2020}\natexlab{}.
\newblock \showarticletitle{{Sound Synthesis, Propagation, and Rendering: {A}
  Survey}}.
\newblock \bibinfo{journal}{\emph{CoRR}}  \bibinfo{volume}{abs/2011.05538}
  (\bibinfo{year}{2020}).
\newblock
\showeprint[arXiv]{2011.05538}
\urldef\tempurl%
\url{https://arxiv.org/abs/2011.05538}
\showURL{%
\tempurl}


\bibitem[\protect\citeauthoryear{Lyon and Laflaquiere}{Lyon and
  Laflaquiere}{2020}]%
        {appleP2020}
\bibfield{author}{\bibinfo{person}{Keith Lyon} {and} \bibinfo{person}{Arnaud
  Laflaquiere}.} \bibinfo{year}{Apple Inc., U.S. Patent 2020/0251882,
  2020}\natexlab{}.
\newblock \bibinfo{title}{{Vertical Emitters with Integral Microlenses}}.
\newblock
\newblock


\bibitem[\protect\citeauthoryear{Mainetti, Patrono, and Sergi}{Mainetti
  et~al\mbox{.}}{2014}]%
        {luca2014}
\bibfield{author}{\bibinfo{person}{Luca Mainetti}, \bibinfo{person}{Luigi
  Patrono}, {and} \bibinfo{person}{Ilaria Sergi}.}
  \bibinfo{year}{2014}\natexlab{}.
\newblock \showarticletitle{{A Survey on Indoor Positioning Systems}}. In
  \bibinfo{booktitle}{\emph{22nd International Conference on Software,
  Telecommunications and Computer Networks (SoftCOM)}}.
  \bibinfo{publisher}{IEEE}, \bibinfo{pages}{111--120}.
\newblock
\urldef\tempurl%
\url{https://doi.org/10.1109/SOFTCOM.2014.7039067}
\showDOI{\tempurl}


\bibitem[\protect\citeauthoryear{Marelli, Aramaki, Kronland-Martinet, and
  Verron}{Marelli et~al\mbox{.}}{2010}]%
        {fire2010}
\bibfield{author}{\bibinfo{person}{Dami\'{a}n Marelli},
  \bibinfo{person}{Mitsuko Aramaki}, \bibinfo{person}{Richard
  Kronland-Martinet}, {and} \bibinfo{person}{Charles Verron}.}
  \bibinfo{year}{2010}\natexlab{}.
\newblock \showarticletitle{{Time-Frequency Synthesis of Noisy Sounds with
  Narrow Spectral Components}}.
\newblock \bibinfo{journal}{\emph{Trans. Audio, Speech and Lang. Proc.}}
  \bibinfo{volume}{18}, \bibinfo{number}{8} (\bibinfo{date}{Nov.}
  \bibinfo{year}{2010}), \bibinfo{pages}{1929–1940}.
\newblock
\showISSN{1063-6676}


\bibitem[\protect\citeauthoryear{McDuffee}{McDuffee}{2014}]%
        {vipe2014}
\bibfield{author}{\bibinfo{person}{Allen McDuffee}.}
  \bibinfo{year}{2014}\natexlab{}.
\newblock \bibinfo{title}{{A Holodeck Videogame Designed to Train Soldiers,
  Jan, 2014. See https://www.wired.com/2014/01/holodeck/}}.
\newblock
\newblock


\bibitem[\protect\citeauthoryear{{Mehra, Ravish and Raghuvanshi, Nikunj and
  Antani, Lakulish and Chandak, Anish and Curtis, Sean and Manocha,
  Dinesh}}{{Mehra, Ravish and Raghuvanshi, Nikunj and Antani, Lakulish and
  Chandak, Anish and Curtis, Sean and Manocha, Dinesh}}{2013}]%
        {mehra2013}
\bibfield{author}{\bibinfo{person}{{Mehra, Ravish and Raghuvanshi, Nikunj and
  Antani, Lakulish and Chandak, Anish and Curtis, Sean and Manocha, Dinesh}}.}
  \bibinfo{year}{2013}\natexlab{}.
\newblock \showarticletitle{Wave-Based Sound Propagation in Large Open Scenes
  Using an Equivalent Source Formulation}.
\newblock \bibinfo{journal}{\emph{ACM Trans. Graph.}} \bibinfo{volume}{32},
  \bibinfo{number}{2}, Article \bibinfo{articleno}{19} (\bibinfo{date}{April}
  \bibinfo{year}{2013}), \bibinfo{numpages}{13}~pages.
\newblock
\showISSN{0730-0301}
\urldef\tempurl%
\url{https://doi.org/10.1145/2451236.2451245}
\showDOI{\tempurl}


\bibitem[\protect\citeauthoryear{Mirri, Prandi, and Salomoni}{Mirri
  et~al\mbox{.}}{2019}]%
        {mirri2019}
\bibfield{author}{\bibinfo{person}{Silvia Mirri}, \bibinfo{person}{Catia
  Prandi}, {and} \bibinfo{person}{Paola Salomoni}.}
  \bibinfo{year}{2019}\natexlab{}.
\newblock \showarticletitle{{Human-Drone Interaction: State of the Art, Open
  Issues and Challenges}}. In \bibinfo{booktitle}{\emph{Proceedings of the ACM
  SIGCOMM 2019 Workshop on Mobile AirGround Edge Computing, Systems, Networks,
  and Applications}} (Beijing, China) \emph{(\bibinfo{series}{MAGESys'19})}.
  \bibinfo{publisher}{Association for Computing Machinery},
  \bibinfo{address}{New York, NY, USA}, \bibinfo{pages}{43–48}.
\newblock
\showISBNx{9781450368797}
\urldef\tempurl%
\url{https://doi.org/10.1145/3341568.3342111}
\showDOI{\tempurl}


\bibitem[\protect\citeauthoryear{Morgan, Chung, and Hadaegh}{Morgan
  et~al\mbox{.}}{2013}]%
        {reactiveColAvMorgan}
\bibfield{author}{\bibinfo{person}{Daniel Morgan}, \bibinfo{person}{Soon-Jo
  Chung}, {and} \bibinfo{person}{Fred Hadaegh}.}
  \bibinfo{year}{2013}\natexlab{}.
\newblock \showarticletitle{{Decentralized Model Predictive Control of Swarms
  of Spacecraft Using Sequential Convex Programming}}. In
  \bibinfo{booktitle}{\emph{Advances in the Astronautical Sciences}}.
\newblock


\bibitem[\protect\citeauthoryear{Morgan, Chung, and Hadaegh}{Morgan
  et~al\mbox{.}}{2014}]%
        {reactiveColAvMorganJ}
\bibfield{author}{\bibinfo{person}{Daniel Morgan}, \bibinfo{person}{Soon-Jo
  Chung}, {and} \bibinfo{person}{Fred~Y. Hadaegh}.}
  \bibinfo{year}{2014}\natexlab{}.
\newblock \showarticletitle{{Model Predictive Control of Swarms of Spacecraft
  Using Sequential Convex Programming}}.
\newblock \bibinfo{journal}{\emph{Journal of Guidance, Control, and Dynamics}}
  \bibinfo{volume}{37}, \bibinfo{number}{6} (\bibinfo{year}{2014}),
  \bibinfo{pages}{1725--1740}.
\newblock
\urldef\tempurl%
\url{https://doi.org/10.2514/1.G000218}
\showDOI{\tempurl}


\bibitem[\protect\citeauthoryear{Mylvaganam, Sassano, and Astolfi}{Mylvaganam
  et~al\mbox{.}}{2017}]%
        {gameCollisionAvoidance2017}
\bibfield{author}{\bibinfo{person}{Thulasi Mylvaganam}, \bibinfo{person}{Mario
  Sassano}, {and} \bibinfo{person}{Alessandro Astolfi}.}
  \bibinfo{year}{2017}\natexlab{}.
\newblock \showarticletitle{{A Differential Game Approach to Multi-Agent
  Collision Avoidance}}.
\newblock \bibinfo{journal}{\emph{IEEE Trans. Automat. Control}}
  \bibinfo{volume}{PP} (\bibinfo{date}{04} \bibinfo{year}{2017}),
  \bibinfo{pages}{4229--4235}.
\newblock
\urldef\tempurl%
\url{https://doi.org/10.1109/TAC.2017.2680602}
\showDOI{\tempurl}


\bibitem[\protect\citeauthoryear{Patterson, Gibson, and Katz}{Patterson
  et~al\mbox{.}}{1988}]%
        {gibson}
\bibfield{author}{\bibinfo{person}{David~A. Patterson}, \bibinfo{person}{Garth
  Gibson}, {and} \bibinfo{person}{Randy~H. Katz}.}
  \bibinfo{year}{1988}\natexlab{}.
\newblock \showarticletitle{{A Case for Redundant Arrays of Inexpensive Disks
  (RAID)}}. In \bibinfo{booktitle}{\emph{Proceedings of the 1988 ACM SIGMOD
  International Conference on Management of Data}} (Chicago, Illinois, USA)
  \emph{(\bibinfo{series}{SIGMOD '88})}. \bibinfo{publisher}{Association for
  Computing Machinery}, \bibinfo{address}{New York, NY, USA},
  \bibinfo{pages}{109–116}.
\newblock
\showISBNx{0897912683}
\urldef\tempurl%
\url{https://doi.org/10.1145/50202.50214}
\showDOI{\tempurl}


\bibitem[\protect\citeauthoryear{Peter}{Peter}{2021}]%
        {motorola2021}
\bibfield{author}{\bibinfo{person}{P. Peter}.} \bibinfo{year}{2021}\natexlab{}.
\newblock \bibinfo{title}{{Motorola Demos Wireless Charger that Works at 3m,
  Handles Four Phones Simultaneously, GSMARENA, September 8, 2021,
  https://www.gsmarena.com/motorola\_demonstrates \_a\_wireless\_charger\_that
  \_works\_at\_up\_to\_3m\_can\_handle\_four\_phones\_simultaneousl-news-50850.php}}.
\newblock
\newblock


\bibitem[\protect\citeauthoryear{Plathottam and Ranganathan}{Plathottam and
  Ranganathan}{2018}]%
        {Plathottam2018NextGD}
\bibfield{author}{\bibinfo{person}{Siby~Jose Plathottam} {and}
  \bibinfo{person}{P. Ranganathan}.} \bibinfo{year}{2018}\natexlab{}.
\newblock \showarticletitle{{Next Generation Distributed and Networked
  Autonomous Vehicles: Review}}.
\newblock \bibinfo{journal}{\emph{2018 10th International Conference on
  Communication Systems \& Networks (COMSNETS)}} (\bibinfo{year}{2018}),
  \bibinfo{pages}{577--582}.
\newblock


\bibitem[\protect\citeauthoryear{Preiss, Hoenig, Ayanian, and Sukhatme}{Preiss
  et~al\mbox{.}}{2017a}]%
        {preiss2017whitewash}
\bibfield{author}{\bibinfo{person}{James Preiss}, \bibinfo{person}{Wolfgang
  Hoenig}, \bibinfo{person}{Nora Ayanian}, {and} \bibinfo{person}{Gaurav
  Sukhatme}.} \bibinfo{year}{2017}\natexlab{a}.
\newblock \showarticletitle{{Downwash-Aware Trajectory Planning for Large
  Quadcopter Teams}}.
\newblock \bibinfo{journal}{\emph{IEEE/RSJ International Conference on
  Intelligent Robots and Systems (IROS)}} (\bibinfo{date}{04}
  \bibinfo{year}{2017}), \bibinfo{pages}{8}.
\newblock


\bibitem[\protect\citeauthoryear{Preiss, Honig, Sukhatme, and Ayanian}{Preiss
  et~al\mbox{.}}{2017b}]%
        {preiss2017}
\bibfield{author}{\bibinfo{person}{James Preiss}, \bibinfo{person}{Wolfgang
  Honig}, \bibinfo{person}{Gaurav Sukhatme}, {and} \bibinfo{person}{Nora
  Ayanian}.} \bibinfo{year}{2017}\natexlab{b}.
\newblock \showarticletitle{{Crazyswarm: A Large Nano-Quadcopter Swarm}}. In
  \bibinfo{booktitle}{\emph{IEEE International Conference on Robotics and
  Automation (ICRA)}}. \bibinfo{pages}{3299--3304}.
\newblock
\urldef\tempurl%
\url{https://doi.org/10.1109/ICRA.2017.7989376}
\showDOI{\tempurl}


\bibitem[\protect\citeauthoryear{Raghuvanshi and Snyder}{Raghuvanshi and
  Snyder}{2014}]%
        {soundParametric2014}
\bibfield{author}{\bibinfo{person}{Nikunj Raghuvanshi} {and}
  \bibinfo{person}{John Snyder}.} \bibinfo{year}{2014}\natexlab{}.
\newblock \showarticletitle{{Parametric Wave Field Coding for Precomputed Sound
  Propagation}}.
\newblock \bibinfo{journal}{\emph{ACM Trans. Graph.}} \bibinfo{volume}{33},
  \bibinfo{number}{4}, Article \bibinfo{articleno}{38} (\bibinfo{date}{July}
  \bibinfo{year}{2014}), \bibinfo{numpages}{11}~pages.
\newblock
\showISSN{0730-0301}
\urldef\tempurl%
\url{https://doi.org/10.1145/2601097.2601184}
\showDOI{\tempurl}


\bibitem[\protect\citeauthoryear{Raghuvanshi and Snyder}{Raghuvanshi and
  Snyder}{2018}]%
        {raghu2018}
\bibfield{author}{\bibinfo{person}{Nikunj Raghuvanshi} {and}
  \bibinfo{person}{John Snyder}.} \bibinfo{year}{2018}\natexlab{}.
\newblock \showarticletitle{{Parametric Directional Coding for Precomputed
  Sound Propagation}}.
\newblock \bibinfo{journal}{\emph{ACM Trans. Graph.}} \bibinfo{volume}{37},
  \bibinfo{number}{4}, Article \bibinfo{articleno}{108} (\bibinfo{date}{July}
  \bibinfo{year}{2018}), \bibinfo{numpages}{14}~pages.
\newblock
\showISSN{0730-0301}
\urldef\tempurl%
\url{https://doi.org/10.1145/3197517.3201339}
\showDOI{\tempurl}


\bibitem[\protect\citeauthoryear{Raspet, L’Espérance, and Daigle}{Raspet
  et~al\mbox{.}}{1995}]%
        {geometric95}
\bibfield{author}{\bibinfo{person}{Richard Raspet}, \bibinfo{person}{André
  L’Espérance}, {and} \bibinfo{person}{Gilles~A. Daigle}.}
  \bibinfo{year}{1995}\natexlab{}.
\newblock \showarticletitle{{The Effect of Realistic Ground Impedance on the
  Accuracy of Ray Tracing}}.
\newblock \bibinfo{journal}{\emph{The Journal of the Acoustical Society of
  America}} \bibinfo{volume}{97}, \bibinfo{number}{1} (\bibinfo{year}{1995}),
  \bibinfo{pages}{154--158}.
\newblock
\urldef\tempurl%
\url{https://doi.org/10.1121/1.412333}
\showDOI{\tempurl}
\showeprint{https://doi.org/10.1121/1.412333}


\bibitem[\protect\citeauthoryear{Ratnasamy, Francis, Handley, Karp, and
  Shenker}{Ratnasamy et~al\mbox{.}}{2001}]%
        {can2001}
\bibfield{author}{\bibinfo{person}{Sylvia Ratnasamy}, \bibinfo{person}{Paul
  Francis}, \bibinfo{person}{Mark Handley}, \bibinfo{person}{Richard Karp},
  {and} \bibinfo{person}{Scott Shenker}.} \bibinfo{year}{2001}\natexlab{}.
\newblock \showarticletitle{{A Scalable Content-Addressable Network}}.
\newblock \bibinfo{journal}{\emph{SIGCOMM Comput. Commun. Rev.}}
  \bibinfo{volume}{31}, \bibinfo{number}{4} (\bibinfo{date}{Aug.}
  \bibinfo{year}{2001}), \bibinfo{pages}{161–172}.
\newblock
\showISSN{0146-4833}
\urldef\tempurl%
\url{https://doi.org/10.1145/964723.383072}
\showDOI{\tempurl}


\bibitem[\protect\citeauthoryear{Ritz, Müller, Hehn, and D'Andrea}{Ritz
  et~al\mbox{.}}{2012}]%
        {opticalpositioning2}
\bibfield{author}{\bibinfo{person}{Robin Ritz}, \bibinfo{person}{Mark~W.
  Müller}, \bibinfo{person}{Markus Hehn}, {and} \bibinfo{person}{Raffaello
  D'Andrea}.} \bibinfo{year}{2012}\natexlab{}.
\newblock \showarticletitle{{Cooperative Quadrocopter Ball Throwing and
  Catching}}. In \bibinfo{booktitle}{\emph{2012 IEEE/RSJ International
  Conference on Intelligent Robots and Systems}}. \bibinfo{pages}{4972--4978}.
\newblock
\urldef\tempurl%
\url{https://doi.org/10.1109/IROS.2012.6385963}
\showDOI{\tempurl}


\bibitem[\protect\citeauthoryear{Rungta, Schissler, Rewkowski, Mehra, and
  Manocha}{Rungta et~al\mbox{.}}{2018}]%
        {hybrid2018}
\bibfield{author}{\bibinfo{person}{A. Rungta}, \bibinfo{person}{C. Schissler},
  \bibinfo{person}{N. Rewkowski}, \bibinfo{person}{R. Mehra}, {and}
  \bibinfo{person}{D. Manocha}.} \bibinfo{year}{2018}\natexlab{}.
\newblock \showarticletitle{{Diffraction Kernels for Interactive Sound
  Propagation in Dynamic Environments}}.
\newblock \bibinfo{journal}{\emph{IEEE Transactions on Visualization \&
  Computer Graphics}} \bibinfo{volume}{24}, \bibinfo{number}{04}
  (\bibinfo{date}{apr} \bibinfo{year}{2018}), \bibinfo{pages}{1613--1622}.
\newblock
\showISSN{1941-0506}
\urldef\tempurl%
\url{https://doi.org/10.1109/TVCG.2018.2794098}
\showDOI{\tempurl}


\bibitem[\protect\citeauthoryear{Schissler, Mehra, and Manocha}{Schissler
  et~al\mbox{.}}{2014}]%
        {diffract2014}
\bibfield{author}{\bibinfo{person}{Carl Schissler}, \bibinfo{person}{Ravish
  Mehra}, {and} \bibinfo{person}{Dinesh Manocha}.}
  \bibinfo{year}{2014}\natexlab{}.
\newblock \showarticletitle{{High-Order Diffraction and Diffuse Reflections for
  Interactive Sound Propagation in Large Environments}}.
\newblock \bibinfo{journal}{\emph{ACM Trans. Graph.}} \bibinfo{volume}{33},
  \bibinfo{number}{4}, Article \bibinfo{articleno}{39} (\bibinfo{date}{July}
  \bibinfo{year}{2014}), \bibinfo{numpages}{12}~pages.
\newblock
\showISSN{0730-0301}
\urldef\tempurl%
\url{https://doi.org/10.1145/2601097.2601216}
\showDOI{\tempurl}


\bibitem[\protect\citeauthoryear{Schulz, Augugliaro, Ritz, and D'Andrea}{Schulz
  et~al\mbox{.}}{2015}]%
        {dandrea2015}
\bibfield{author}{\bibinfo{person}{Maximilian Schulz},
  \bibinfo{person}{Federico Augugliaro}, \bibinfo{person}{Robin Ritz}, {and}
  \bibinfo{person}{Raffaello D'Andrea}.} \bibinfo{year}{2015}\natexlab{}.
\newblock \showarticletitle{{High-speed, Steady Flight with a Quadrocopter in a
  Confined Environment using a Tether}}. In \bibinfo{booktitle}{\emph{2015
  IEEE/RSJ International Conference on Intelligent Robots and Systems (IROS)}}.
  \bibinfo{pages}{1279--1284}.
\newblock
\urldef\tempurl%
\url{https://doi.org/10.1109/IROS.2015.7353533}
\showDOI{\tempurl}


\bibitem[\protect\citeauthoryear{Shakhatreh, Sawalmeh, Al-Fuqaha, Dou, Almaita,
  Khalil, Othman, Khreishah, and Guizani}{Shakhatreh et~al\mbox{.}}{2019}]%
        {Shakhatreh2019UnmannedAV}
\bibfield{author}{\bibinfo{person}{Hazim Shakhatreh}, \bibinfo{person}{A.
  Sawalmeh}, \bibinfo{person}{Ala Al-Fuqaha}, \bibinfo{person}{Zuochao Dou},
  \bibinfo{person}{Eyad~K. Almaita}, \bibinfo{person}{Issa~M. Khalil},
  \bibinfo{person}{Noor~Shamsiah Othman}, \bibinfo{person}{A. Khreishah}, {and}
  \bibinfo{person}{M. Guizani}.} \bibinfo{year}{2019}\natexlab{}.
\newblock \showarticletitle{{Unmanned Aerial Vehicles (UAVs): A Survey on Civil
  Applications and Key Research Challenges}}.
\newblock \bibinfo{journal}{\emph{IEEE Access}}  \bibinfo{volume}{7}
  (\bibinfo{year}{2019}), \bibinfo{pages}{48572--48634}.
\newblock


\bibitem[\protect\citeauthoryear{Shivgan and Dong}{Shivgan and Dong}{2020}]%
        {pathplanning2020}
\bibfield{author}{\bibinfo{person}{Rutuja Shivgan} {and}
  \bibinfo{person}{Ziqian Dong}.} \bibinfo{year}{2020}\natexlab{}.
\newblock \showarticletitle{{Energy-Efficient Drone Coverage Path Planning
  using Genetic Algorithm}}. In \bibinfo{booktitle}{\emph{IEEE 21st
  International Conference on High Performance Switching and Routing (HPSR)}}.
  \bibinfo{publisher}{IEEE}, \bibinfo{pages}{1--6}.
\newblock
\urldef\tempurl%
\url{https://doi.org/10.1109/HPSR48589.2020.9098989}
\showDOI{\tempurl}


\bibitem[\protect\citeauthoryear{Shpunt, Medioni, Cohen, Szli, and
  Deitch}{Shpunt et~al\mbox{.}}{2017}]%
        {appleP2017}
\bibfield{author}{\bibinfo{person}{Alexander Shpunt}, \bibinfo{person}{Gerard
  Medioni}, \bibinfo{person}{Daniel Cohen}, \bibinfo{person}{Erez Szli}, {and}
  \bibinfo{person}{Ronen Deitch}.} \bibinfo{year}{Apple Inc., U.S. Patent
  9,582,889 B2, 2017}\natexlab{}.
\newblock \bibinfo{title}{{Depth Mapping Based on Pattern Matching and
  Stereoscopic Information}}.
\newblock
\newblock


\bibitem[\protect\citeauthoryear{Stephan, Alonso-Mora, and Siegwart}{Stephan
  et~al\mbox{.}}{2013}]%
        {reactiveColAvoidance2013}
\bibfield{author}{\bibinfo{person}{Martin Stephan}, \bibinfo{person}{Javier
  Alonso-Mora}, {and} \bibinfo{person}{Roland Siegwart}.}
  \bibinfo{year}{2013}\natexlab{}.
\newblock \showarticletitle{{Reciprocal Collision Avoidance With Motion
  Continuity Constraints}}.
\newblock \bibinfo{journal}{\emph{IEEE Transactions on Robotics}}
  \bibinfo{volume}{29} (\bibinfo{date}{08} \bibinfo{year}{2013}),
  \bibinfo{pages}{899--912}.
\newblock
\urldef\tempurl%
\url{https://doi.org/10.1109/TRO.2013.2258733}
\showDOI{\tempurl}


\bibitem[\protect\citeauthoryear{Stoica, Morris, Karger, Kaashoek, and
  Balakrishnan}{Stoica et~al\mbox{.}}{2001}]%
        {chord2001}
\bibfield{author}{\bibinfo{person}{Ion Stoica}, \bibinfo{person}{Robert
  Morris}, \bibinfo{person}{David Karger}, \bibinfo{person}{M.~Frans Kaashoek},
  {and} \bibinfo{person}{Hari Balakrishnan}.} \bibinfo{year}{2001}\natexlab{}.
\newblock \showarticletitle{{Chord: A Scalable Peer-to-Peer Lookup Service for
  Internet Applications}}.
\newblock \bibinfo{journal}{\emph{SIGCOMM Comput. Commun. Rev.}}
  \bibinfo{volume}{31}, \bibinfo{number}{4} (\bibinfo{date}{Aug.}
  \bibinfo{year}{2001}), \bibinfo{pages}{149–160}.
\newblock
\showISSN{0146-4833}
\urldef\tempurl%
\url{https://doi.org/10.1145/964723.383071}
\showDOI{\tempurl}


\bibitem[\protect\citeauthoryear{Sun, Qi, Wu, and Wang}{Sun
  et~al\mbox{.}}{2020}]%
        {Sun2020PathPF}
\bibfield{author}{\bibinfo{person}{Hang Sun}, \bibinfo{person}{Juntong Qi},
  \bibinfo{person}{Chong Wu}, {and} \bibinfo{person}{Mingming Wang}.}
  \bibinfo{year}{2020}\natexlab{}.
\newblock \showarticletitle{{Path Planning for Dense Drone Formation Based on
  Modified Artificial Potential Fields}}.
\newblock \bibinfo{journal}{\emph{39th Chinese Control Conference (CCC)}}
  (\bibinfo{year}{2020}), \bibinfo{pages}{4658--4664}.
\newblock


\bibitem[\protect\citeauthoryear{Sun, Tang, and Lao}{Sun et~al\mbox{.}}{2017}]%
        {repel1}
\bibfield{author}{\bibinfo{person}{Jiayi Sun}, \bibinfo{person}{Jun Tang},
  {and} \bibinfo{person}{Songyang Lao}.} \bibinfo{year}{2017}\natexlab{}.
\newblock \showarticletitle{{Collision Avoidance for Cooperative UAVs With
  Optimized Artificial Potential Field Algorithm}}.
\newblock \bibinfo{journal}{\emph{IEEE Access}}  \bibinfo{volume}{PP}
  (\bibinfo{date}{08} \bibinfo{year}{2017}), \bibinfo{pages}{18382--18390}.
\newblock
\urldef\tempurl%
\url{https://doi.org/10.1109/ACCESS.2017.2746752}
\showDOI{\tempurl}


\bibitem[\protect\citeauthoryear{Sutherland}{Sutherland}{1968}]%
        {Sutherland1968AHT}
\bibfield{author}{\bibinfo{person}{Ivan~E. Sutherland}.}
  \bibinfo{year}{1968}\natexlab{}.
\newblock \showarticletitle{{A Head-Mounted Three Dimensional Display}}. In
  \bibinfo{booktitle}{\emph{AFIPS '68 (Fall, part I)}}.
\newblock


\bibitem[\protect\citeauthoryear{Tabasso, Cichella, Mehdi, Marinho, and
  Hovakimyan}{Tabasso et~al\mbox{.}}{2021}]%
        {speedAdjust2021}
\bibfield{author}{\bibinfo{person}{Camilla Tabasso}, \bibinfo{person}{Venanzio
  Cichella}, \bibinfo{person}{Syed Mehdi}, \bibinfo{person}{Thiago Marinho},
  {and} \bibinfo{person}{Naira Hovakimyan}.} \bibinfo{year}{2021}\natexlab{}.
\newblock \showarticletitle{{Time Coordination and Collision Avoidance Using
  Leader-Follower Strategies in Multi-Vehicle Missions}}.
\newblock \bibinfo{journal}{\emph{Robotics}}  \bibinfo{volume}{10}
  (\bibinfo{date}{02} \bibinfo{year}{2021}), \bibinfo{pages}{34}.
\newblock
\urldef\tempurl%
\url{https://doi.org/10.3390/robotics10010034}
\showDOI{\tempurl}


\bibitem[\protect\citeauthoryear{Tang, Meng, and Manocha}{Tang
  et~al\mbox{.}}{2021}]%
        {soundLearn2020}
\bibfield{author}{\bibinfo{person}{Zhenyu Tang}, \bibinfo{person}{Hsien{-}Yu
  Meng}, {and} \bibinfo{person}{Dinesh Manocha}.}
  \bibinfo{year}{2021}\natexlab{}.
\newblock \showarticletitle{{Learning Acoustic Scattering Fields for Highly
  Dynamic Interactive Sound Propagation}}. In \bibinfo{booktitle}{\emph{IEEE
  Virtual Reality and 3D User Interfaces (VR)}}. \bibinfo{publisher}{IEEE},
  \bibinfo{pages}{826--836}.
\newblock


\bibitem[\protect\citeauthoryear{Tsingos, Funkhouser, Ngan, and
  Carlbom}{Tsingos et~al\mbox{.}}{2001}]%
        {diffract2001}
\bibfield{author}{\bibinfo{person}{Nicolas Tsingos}, \bibinfo{person}{Thomas
  Funkhouser}, \bibinfo{person}{Addy Ngan}, {and} \bibinfo{person}{Ingrid
  Carlbom}.} \bibinfo{year}{2001}\natexlab{}.
\newblock \showarticletitle{{Modeling Acoustics in Virtual Environments Using
  the Uniform Theory of Diffraction}}. In \bibinfo{booktitle}{\emph{Proceedings
  of the 28th Annual Conference on Computer Graphics and Interactive
  Techniques}} \emph{(\bibinfo{series}{SIGGRAPH '01})}.
  \bibinfo{publisher}{Association for Computing Machinery},
  \bibinfo{address}{New York, NY, USA}, \bibinfo{pages}{545–552}.
\newblock
\showISBNx{158113374X}
\urldef\tempurl%
\url{https://doi.org/10.1145/383259.383323}
\showDOI{\tempurl}


\bibitem[\protect\citeauthoryear{Tsingos, Gallo, and Drettakis}{Tsingos
  et~al\mbox{.}}{2004}]%
        {rendering2004}
\bibfield{author}{\bibinfo{person}{Nicolas Tsingos}, \bibinfo{person}{Emmanuel
  Gallo}, {and} \bibinfo{person}{George Drettakis}.}
  \bibinfo{year}{2004}\natexlab{}.
\newblock \showarticletitle{{Perceptual Audio Rendering of Complex Virtual
  Environments}}.
\newblock \bibinfo{journal}{\emph{ACM Trans. Graph.}} \bibinfo{volume}{23},
  \bibinfo{number}{3} (\bibinfo{date}{Aug.} \bibinfo{year}{2004}),
  \bibinfo{pages}{249–258}.
\newblock
\showISSN{0730-0301}
\urldef\tempurl%
\url{https://doi.org/10.1145/1015706.1015710}
\showDOI{\tempurl}


\bibitem[\protect\citeauthoryear{Umetani, Panotopoulou, Schmidt, and
  Whiting}{Umetani et~al\mbox{.}}{2016}]%
        {printone2016}
\bibfield{author}{\bibinfo{person}{Nobuyuki Umetani}, \bibinfo{person}{Athina
  Panotopoulou}, \bibinfo{person}{Ryan Schmidt}, {and} \bibinfo{person}{Emily
  Whiting}.} \bibinfo{year}{2016}\natexlab{}.
\newblock \showarticletitle{{Printone: Interactive Resonance Simulation for
  Free-Form Print-Wind Instrument Design}}.
\newblock \bibinfo{journal}{\emph{ACM Trans. Graph.}} \bibinfo{volume}{35},
  \bibinfo{number}{6}, Article \bibinfo{articleno}{184} (\bibinfo{date}{Nov.}
  \bibinfo{year}{2016}), \bibinfo{numpages}{14}~pages.
\newblock
\showISSN{0730-0301}
\urldef\tempurl%
\url{https://doi.org/10.1145/2980179.2980250}
\showDOI{\tempurl}


\bibitem[\protect\citeauthoryear{van~den Berg, Guy, Lin, and Manocha}{van~den
  Berg et~al\mbox{.}}{2011a}]%
        {ReactiveCollisionAvoidance2011}
\bibfield{author}{\bibinfo{person}{Jur van~den Berg}, \bibinfo{person}{Stephen
  Guy}, \bibinfo{person}{Ming Lin}, {and} \bibinfo{person}{Dinesh Manocha}.}
  \bibinfo{year}{2011}\natexlab{a}.
\newblock \bibinfo{booktitle}{\emph{{Reciprocal n-Body Collision Avoidance}}}.
  Vol.~\bibinfo{volume}{70}.
\newblock \bibinfo{pages}{3--19}.
\newblock
\showISBNx{978-3-642-19456-6}
\urldef\tempurl%
\url{https://doi.org/10.1007/978-3-642-19457-3_1}
\showDOI{\tempurl}


\bibitem[\protect\citeauthoryear{van~den Berg, Lin, and Manocha}{van~den Berg
  et~al\mbox{.}}{2008}]%
        {ReactiveCollisionAvoidance2008}
\bibfield{author}{\bibinfo{person}{Jur van~den Berg}, \bibinfo{person}{Ming
  Lin}, {and} \bibinfo{person}{Dinesh Manocha}.}
  \bibinfo{year}{2008}\natexlab{}.
\newblock \showarticletitle{{Reciprocal Velocity Obstacles for Real-Time
  Multi-agent Navigation}}.
\newblock \bibinfo{journal}{\emph{ICRA}}, \bibinfo{pages}{1928--1935}.
\newblock
\urldef\tempurl%
\url{https://doi.org/10.1109/ROBOT.2008.4543489}
\showDOI{\tempurl}


\bibitem[\protect\citeauthoryear{van~den Berg, Snape, Guy, and Manocha}{van~den
  Berg et~al\mbox{.}}{2011b}]%
        {ReactiveCollisionAvoidance20112}
\bibfield{author}{\bibinfo{person}{Jur van~den Berg}, \bibinfo{person}{Jamie
  Snape}, \bibinfo{person}{Stephen Guy}, {and} \bibinfo{person}{Dinesh
  Manocha}.} \bibinfo{year}{2011}\natexlab{b}.
\newblock \showarticletitle{{Reciprocal Collision Avoidance with
  Acceleration-Velocity Obstacles}}. \bibinfo{pages}{3475--3482}.
\newblock
\urldef\tempurl%
\url{https://doi.org/10.1109/ICRA.2011.5980408}
\showDOI{\tempurl}


\bibitem[\protect\citeauthoryear{van~den Berg, Wilkie, Guy, Niethammer, and
  Manocha}{van~den Berg et~al\mbox{.}}{2012}]%
        {downwash1}
\bibfield{author}{\bibinfo{person}{Jur van~den Berg}, \bibinfo{person}{David
  Wilkie}, \bibinfo{person}{Stephen Guy}, \bibinfo{person}{Marc Niethammer},
  {and} \bibinfo{person}{Dinesh Manocha}.} \bibinfo{year}{2012}\natexlab{}.
\newblock \showarticletitle{{LQG-obstacles: Feedback Control with Collision
  Avoidance for Mobile Robots with Motion and Sensing Uncertainty}}.
\newblock \bibinfo{journal}{\emph{Proceedings - IEEE International Conference
  on Robotics and Automation}} (\bibinfo{date}{05} \bibinfo{year}{2012}),
  \bibinfo{pages}{346--353}.
\newblock
\urldef\tempurl%
\url{https://doi.org/10.1109/ICRA.2012.6224648}
\showDOI{\tempurl}


\bibitem[\protect\citeauthoryear{Vlaardinger and Broek}{Vlaardinger and
  Broek}{1999}]%
        {mri1999}
\bibfield{author}{\bibinfo{person}{Marinus~T. Vlaardinger} {and}
  \bibinfo{person}{Jacques~A. Broek}.} \bibinfo{year}{1999}\natexlab{}.
\newblock \bibinfo{booktitle}{\emph{{Magnetic Resonance Imaging Theory and
  Practice}} (\bibinfo{edition}{2} ed.)}.
\newblock \bibinfo{publisher}{Springer-Verlag Berlin Heidelberg}.
\newblock
\showISBNx{978-3-662-03800-0}
\urldef\tempurl%
\url{https://doi.org/10.1007/978-3-662-03800-0}
\showDOI{\tempurl}


\bibitem[\protect\citeauthoryear{Wand and Stra\ss{}er}{Wand and
  Stra\ss{}er}{2004}]%
        {soundRendering2004}
\bibfield{author}{\bibinfo{person}{M. Wand} {and} \bibinfo{person}{W.
  Stra\ss{}er}.} \bibinfo{year}{2004}\natexlab{}.
\newblock \showarticletitle{{Multi-Resolution Sound Rendering}}. In
  \bibinfo{booktitle}{\emph{Proceedings of the First Eurographics Conference on
  Point-Based Graphics}} (Switzerland) \emph{(\bibinfo{series}{SPBG'04})}.
  \bibinfo{publisher}{Eurographics Association}, \bibinfo{address}{Goslar,
  DEU}, \bibinfo{pages}{3–11}.
\newblock
\showISBNx{3905673096}


\bibitem[\protect\citeauthoryear{Wieringa, Boum, Eendebak, van Basten,
  Beerlage, Smits, and Bos}{Wieringa et~al\mbox{.}}{2004}]%
        {davinci2014}
\bibfield{author}{\bibinfo{person}{Fokko~P Wieringa}, \bibinfo{person}{Henri
  Boum}, \bibinfo{person}{Pieter~T Eendebak}, \bibinfo{person}{Jean-Paul~A van
  Basten}, \bibinfo{person}{Harrie~P Beerlage}, \bibinfo{person}{Geert~A.
  Smits}, {and} \bibinfo{person}{Jelte~E Bos}.}
  \bibinfo{year}{2004}\natexlab{}.
\newblock \showarticletitle{{Improved Depth Perception with Three-dimensional
  Auxiliary Display and Computer Generated Three-dimensional Panoramic
  Overviews in Robot-assisted Laparoscopy}}.
\newblock \bibinfo{journal}{\emph{Journal of Medical Imaging (Bellingham,
  Wash.)}} \bibinfo{volume}{1}, \bibinfo{number}{1} (\bibinfo{date}{April}
  \bibinfo{year}{2004}).
\newblock


\bibitem[\protect\citeauthoryear{Williams}{Williams}{2018}]%
        {nytimes2018}
\bibfield{author}{\bibinfo{person}{Stephen Williams}.}
  \bibinfo{year}{2018}\natexlab{}.
\newblock \bibinfo{title}{{For Electric Cars Without a Plug, Thank Tesla (the
  Scientist), The New York Times, May 31, 2018,
  https://www.nytimes.com/2018/05/31/business/electric-cars-wireless-charging.html}}.
\newblock
\newblock


\bibitem[\protect\citeauthoryear{Wu, Low, Pang, and Tan}{Wu
  et~al\mbox{.}}{2021}]%
        {urbanplanning2021}
\bibfield{author}{\bibinfo{person}{Yu Wu}, \bibinfo{person}{Kin~Huat Low},
  \bibinfo{person}{Bizhao Pang}, {and} \bibinfo{person}{Qingyu Tan}.}
  \bibinfo{year}{2021}\natexlab{}.
\newblock \showarticletitle{{Swarm-Based 4D Path Planning For Drone Operations
  in Urban Environments}}.
\newblock \bibinfo{journal}{\emph{IEEE Transactions on Vehicular Technology}}
  \bibinfo{volume}{70}, \bibinfo{number}{8} (\bibinfo{year}{2021}),
  \bibinfo{pages}{7464--7479}.
\newblock
\urldef\tempurl%
\url{https://doi.org/10.1109/TVT.2021.3093318}
\showDOI{\tempurl}


\bibitem[\protect\citeauthoryear{Wu, Song, Khosla, Zhang, Tang, and Xiao}{Wu
  et~al\mbox{.}}{2015}]%
        {3DdeepLearning2015}
\bibfield{author}{\bibinfo{person}{Zhirong Wu}, \bibinfo{person}{Shuran Song},
  \bibinfo{person}{Aditya Khosla}, \bibinfo{person}{Linguang Zhang},
  \bibinfo{person}{Xiaoou Tang}, {and} \bibinfo{person}{Jianxiong Xiao}.}
  \bibinfo{year}{2015}\natexlab{}.
\newblock \showarticletitle{{3D ShapeNets: A Deep Representation for Volumetric
  Shape Modeling}}. In \bibinfo{booktitle}{\emph{IEEE Conference on Computer
  Vision and Pattern Recognition (CVPR)}}. \bibinfo{publisher}{IEEE},
  \bibinfo{address}{Boston, USA}, \bibinfo{pages}{1912--1920}.
\newblock


\bibitem[\protect\citeauthoryear{Xu and Liu}{Xu and Liu}{2020}]%
        {Xu2020PhysicsGuidedSS}
\bibfield{author}{\bibinfo{person}{Siqi Xu} {and} \bibinfo{person}{Shiguang
  Liu}.} \bibinfo{year}{2020}\natexlab{}.
\newblock \showarticletitle{{Physics-Guided Sound Synthesis for Rotating
  Blades}}. In \bibinfo{booktitle}{\emph{Advances in Computer Graphics}}.
  \bibinfo{pages}{233--244}.
\newblock
\urldef\tempurl%
\url{https://doi.org/10.1007/978-3-030-61864-3_20}
\showDOI{\tempurl}


\bibitem[\protect\citeauthoryear{Xu, Lai, Li, Luo, and You}{Xu
  et~al\mbox{.}}{2019}]%
        {planning2019}
\bibfield{author}{\bibinfo{person}{Yang Xu}, \bibinfo{person}{Shupeng Lai},
  \bibinfo{person}{Jiaxin Li}, \bibinfo{person}{Delin Luo}, {and}
  \bibinfo{person}{Yancheng You}.} \bibinfo{year}{2019}\natexlab{}.
\newblock \showarticletitle{{Concurrent Optimal Trajectory Planning for Indoor
  Quadrotor Formation Switching}}.
\newblock \bibinfo{journal}{\emph{Journal of Intelligent \& Robotic Systems}}
  \bibinfo{volume}{94} (\bibinfo{date}{05} \bibinfo{year}{2019}).
\newblock
\urldef\tempurl%
\url{https://doi.org/10.1007/s10846-018-0813-9}
\showDOI{\tempurl}


\bibitem[\protect\citeauthoryear{Yan, Yang, Yumer, Guo, and Lee}{Yan
  et~al\mbox{.}}{2016}]%
        {nipsLearning3Dobjects}
\bibfield{author}{\bibinfo{person}{Xinchen Yan}, \bibinfo{person}{Jimei Yang},
  \bibinfo{person}{Ersin Yumer}, \bibinfo{person}{Yijie Guo}, {and}
  \bibinfo{person}{Honglak Lee}.} \bibinfo{year}{2016}\natexlab{}.
\newblock \showarticletitle{{Perspective Transformer Nets: Learning Single-View
  3D Object Reconstruction without 3D Supervision}}. In
  \bibinfo{booktitle}{\emph{Proceedings of the 30th International Conference on
  Neural Information Processing Systems}} (Barcelona, Spain)
  \emph{(\bibinfo{series}{NIPS'16})}. \bibinfo{publisher}{Curran Associates
  Inc.}, \bibinfo{address}{Red Hook, NY, USA}, \bibinfo{pages}{1704–1712}.
\newblock
\showISBNx{9781510838819}


\bibitem[\protect\citeauthoryear{Yeh, Mehra, Ren, Antani, Manocha, and Lin}{Yeh
  et~al\mbox{.}}{2013}]%
        {hybrid2013}
\bibfield{author}{\bibinfo{person}{Hengchin Yeh}, \bibinfo{person}{Ravish
  Mehra}, \bibinfo{person}{Zhimin Ren}, \bibinfo{person}{Lakulish Antani},
  \bibinfo{person}{Dinesh Manocha}, {and} \bibinfo{person}{Ming Lin}.}
  \bibinfo{year}{2013}\natexlab{}.
\newblock \showarticletitle{{Wave-Ray Coupling for Interactive Sound
  Propagation in Large Complex Scenes}}.
\newblock \bibinfo{journal}{\emph{ACM Trans. Graph.}} \bibinfo{volume}{32},
  \bibinfo{number}{6}, Article \bibinfo{articleno}{165} (\bibinfo{date}{Nov.}
  \bibinfo{year}{2013}), \bibinfo{numpages}{11}~pages.
\newblock
\showISSN{0730-0301}
\urldef\tempurl%
\url{https://doi.org/10.1145/2508363.2508420}
\showDOI{\tempurl}


\bibitem[\protect\citeauthoryear{You and Jiang}{You and Jiang}{2020}]%
        {quanzeng2020}
\bibfield{author}{\bibinfo{person}{Quanzeng You} {and} \bibinfo{person}{Hao
  Jiang}.} \bibinfo{year}{2020}\natexlab{}.
\newblock \showarticletitle{{Real-time 3D Deep Multi-Camera Tracking}}.
\newblock \bibinfo{journal}{\emph{CoRR}}  \bibinfo{volume}{abs/2003.11753}
  (\bibinfo{year}{2020}), \bibinfo{numpages}{17}~pages.
\newblock
\showeprint[arXiv]{2003.11753}
\urldef\tempurl%
\url{https://arxiv.org/abs/2003.11753}
\showURL{%
\tempurl}


\bibitem[\protect\citeauthoryear{Zabatani, Surazhsky, Sperling, Moshe, Menashe,
  Silver, Karni, Bronstein, Bronstein, and Kimmel}{Zabatani
  et~al\mbox{.}}{2020}]%
        {realsense2020}
\bibfield{author}{\bibinfo{person}{Aviad Zabatani}, \bibinfo{person}{Vitaly
  Surazhsky}, \bibinfo{person}{Erez Sperling}, \bibinfo{person}{Sagi~Ben
  Moshe}, \bibinfo{person}{Ohad Menashe}, \bibinfo{person}{David~H. Silver},
  \bibinfo{person}{Zachi Karni}, \bibinfo{person}{Alexander~M. Bronstein},
  \bibinfo{person}{Michael~M. Bronstein}, {and} \bibinfo{person}{Ron Kimmel}.}
  \bibinfo{year}{2020}\natexlab{}.
\newblock \showarticletitle{Intel® RealSense™ SR300 Coded Light Depth
  Camera}.
\newblock \bibinfo{journal}{\emph{IEEE Transactions on Pattern Analysis and
  Machine Intelligence}} \bibinfo{volume}{42}, \bibinfo{number}{10}
  (\bibinfo{year}{2020}), \bibinfo{pages}{2333--2345}.
\newblock
\urldef\tempurl%
\url{https://doi.org/10.1109/TPAMI.2019.2915841}
\showDOI{\tempurl}


\bibitem[\protect\citeauthoryear{Zafari, Gkelias, and Leung}{Zafari
  et~al\mbox{.}}{2019}]%
        {zafari2019}
\bibfield{author}{\bibinfo{person}{Faheem Zafari}, \bibinfo{person}{Athanasios
  Gkelias}, {and} \bibinfo{person}{Kin~K. Leung}.}
  \bibinfo{year}{2019}\natexlab{}.
\newblock \showarticletitle{{A Survey of Indoor Localization Systems and
  Technologies}}.
\newblock \bibinfo{journal}{\emph{IEEE Communications Surveys and Tutorials}}
  \bibinfo{volume}{21}, \bibinfo{number}{3} (\bibinfo{year}{2019}),
  \bibinfo{pages}{2568--2599}.
\newblock
\urldef\tempurl%
\url{https://doi.org/10.1109/COMST.2019.2911558}
\showDOI{\tempurl}


\bibitem[\protect\citeauthoryear{Zhan}{Zhan}{2021}]%
        {guinessWorldRecord}
\bibfield{author}{\bibinfo{person}{Echo Zhan}.}
  \bibinfo{year}{2021}\natexlab{}.
\newblock \bibinfo{title}{{3,281 Drones Break Dazzling Record for Most Airborne
  Simultaneously, May 17, 2021. See https://www.guinnessworldrecords.com/news/
  commercial/2021/5/3281-drones-break-dazzling-record-for-most-airborne-simultaneously-655062}}.
\newblock
\newblock


\bibitem[\protect\citeauthoryear{Zhou, Wang, Fang, Bui, and Berg}{Zhou
  et~al\mbox{.}}{2017}]%
        {wild2017}
\bibfield{author}{\bibinfo{person}{Yipin Zhou}, \bibinfo{person}{Zhaowen Wang},
  \bibinfo{person}{Chen Fang}, \bibinfo{person}{Trung Bui}, {and}
  \bibinfo{person}{Tamara Berg}.} \bibinfo{year}{2017}\natexlab{}.
\newblock \showarticletitle{{Visual to Sound: Generating Natural Sound for
  Videos in the Wild}}. In \bibinfo{booktitle}{\emph{Proceedings of the IEEE
  Conference on Computer Vision and Pattern Recognition}}.
  \bibinfo{pages}{3550--3558}.
\newblock


\end{thebibliography}

\end{document}